\documentclass[english,12pt]{article}
\usepackage[T1]{fontenc}
\usepackage[utf8]{inputenc}
\usepackage[english]{babel}
\usepackage[pdftex]{graphicx}
\usepackage[width=\textwidth]{caption}
\usepackage{amsmath,amsopn,amssymb,bm, dsfont,mathrsfs}
\usepackage{a4wide}
\usepackage{array,float, placeins,flafter, longtable,booktabs,pdflscape}
\usepackage{verbatim}
\usepackage{color}
\usepackage{natbib}
\usepackage{lmodern}
\usepackage{xr}
\usepackage{mathtools}
\usepackage[shortlabels]{enumitem}
\usepackage{relsize}

\makeatletter

\newtheorem{thm}{Theorem}[section]
\newtheorem{cor}[thm]{Corollary}
\newtheorem{lem}{Lemma}[section]
\newtheorem{prop}[thm]{Proposition}

\newtheorem{definition}[thm]{Definition}
\newtheorem{example}{Example}
\newtheorem{hyp}{Assumption}
\numberwithin{equation}{section}

\newcommand{\eps}{\varepsilon}

\newcommand{\indep}{\perp \!\!\! \perp}

\definecolor{darkgreen}{rgb}{0.1,0.7,0.2}

\renewcommand{\citep}[1]{\citeauthor{#1}, \citeyear{#1}}

\newcommand{\argmin}{\text{argmin }}
\newcommand{\st}[1]{\texttt{#1}}

\renewcommand{\section}{\@startsection{section}{2}{0mm}{-1.5\baselineskip}{1\baselineskip}{\normalfont\large\bfseries}}
\renewcommand{\subsection}{\@startsection{subsection}{2}{0mm}{-1.2\baselineskip}{1\baselineskip}{\normalfont\normalsize\bfseries}}
\renewcommand{\subsubsection}{\@startsection{subsubsection}{3}{0mm}{-0.8\baselineskip}{0.4\baselineskip}{\normalfont\normalsize\itshape}}

\date{First version: May 18th, 2021 \\ This version: \today}

\begin{document}

\title{Trading-off Bias and Variance When the Size of the Treatment Effect is Bounded\thanks{This paper was previously circulated under the following title: ``The Minimax Estimator of the Average Treatment Effect, among Linear Combinations of Estimators of Bounded Conditional Average Treatment Effects''. I am extremely grateful to Henri Fabre and Luca Moreno-Louzada for their outstanding
research assistance. I am also very grateful to Timothy Armstrong, Xavier D'Haultf\oe{}uille, Michal Koles\'{a}r, and seminar participants at the Toulouse School of Economics for their helpful comments. I was funded by the European Union (ERC, REALLYCREDIBLE, GA Number
101043899). Views and opinions expressed are those of the authors and do not reflect those of the European Union
or the European Research Council Executive Agency. Neither the European Union nor the granting authority
can be held responsible.}}

\author{Cl\'{e}ment de Chaisemartin\thanks{de Chaisemartin: Sciences Po, Economics Department (email: clement.dechaisemartin@sciencespo.fr)}}

\maketitle ~\vspace{-1cm}

\begin{abstract}
Assume that one is interested in estimating an average treatment effect (ATE), equal to a weighted average of $S$ conditional average treatment effects (CATEs). One has unbiased estimators of the CATEs.  One could just average the CATE estimators, to form an unbiased estimator of the ATE. However, some CATE estimators may be less precise than others. Then, downweighting the imprecisely estimated CATEs may lead to a lower mean-squared error and/or shorter confidence intervals. This paper investigates this bias-variance trade-off, by deriving minimax-linear estimators of, and confidence intervals (CI) for, the ATE, under various restrictions on the CATEs. First, I assume that the magnitude of the CATEs is bounded. Then I assume that their heterogeneity is bounded. I use my results to revisit two empirical applications, and find that minimax-linear estimators and CIs lead to small but non-negligible precision gains, of around 5\%.
\end{abstract}
\textbf{Keywords:} stratified randomized controlled trial, matching estimator, trimming, bias-variance trade-off, average treatment effect, minimax-linear estimator, minimax-length confidence intervals, shrinkage.

\medskip
\textbf{JEL Codes:} C21, C23

\section{Introduction}

Assume one is interested in estimating an average treatment effect (ATE) $\tau$, equal to a weighted average of $S$ conditional average treatment effects (CATEs) $(\tau_s)_{1\leq s\leq S}$:
\begin{align}\label{eq:decompo_intro}
    &\tau =  \sum_{s=1}^Sp_s\tau_s,
\end{align}
where $(p_s)_{1\leq s\leq S}$ are known positive weights that sum to 1.
Moreover, one has unbiased estimators $(\widehat{\tau}_s)_{1\leq s\leq S}$ of the CATEs:
\begin{align}\label{eq:decompo_intro2}
   &E(\widehat{\tau}_s)=\tau_s.
\end{align}
\eqref{eq:decompo_intro} and \eqref{eq:decompo_intro2} for instance hold in stratified randomized controlled trials (SRCTs). There, $s$ indexes strata, $p_s$ is the proportion $s$ accounts for in the population, $\tau_s$ is the CATE in stratum $s$, $\tau$ is the ATE, and $\widehat{\tau}_s$ compares the average outcome of treated and untreated units in stratum $s$. As explained below, this setting also applies to matching studies. 

\medskip
When \eqref{eq:decompo_intro} and \eqref{eq:decompo_intro2} hold,
to estimate $\tau$, one can just use the unbiased estimator
$$\widehat{\tau}(\bm{p}):=\sum_{s=1}^Sp_s\widehat{\tau}_s.$$ However, some CATE estimators may be less precise than others. For instance, this will arise in an SRCT, if the treatment probability is far from 1/2 in some strata and close to 1/2 in other strata. Then, downweighting the imprecisely estimated CATEs may lead to a lower mean-squared error. This paper investigates this bias-variance trade-off, by deriving minimax-linear estimators of, and confidence intervals (CI) for, $\tau$. Whenever possible, I derive closed-form or quasi-closed-form expressions of those estimators and CIs, in an effort to make them more transparent.

\medskip
If CATEs are unbounded, downweighting may lead to an unbounded bias, that will dominate any decrease in variance. To allow for non-trivial bias-variance trade-offs, one needs to impose restrictions on the CATEs. First, I assume that their magnitude is bounded: for all $s$, $|\tau_s|\leq B$ for a known $B$. This restriction is appealing, as applied researchers often have a good ex-ante sense of what would be an implausibly large effect for the intervention they consider. That prior might be based on the available literature studying related interventions. For instance, my first empirical application considers the effect, on students' test scores, of a school with a pedagogy and organization fairly close to so-called ``No Excuse'' charter schools. In a review of four papers estimating 22 CATEs for such schools, none of the 22 estimated CATEs is larger than 35\% of a standard deviation, and for 20 of the 22 CATEs we can reject at the 90\% level the null that the CATE is larger than 50\%. Based on these prior studies, applied researchers may find it plausible to assume ex-ante that the CATEs of the intervention I consider cannot be larger than 50\% of a standard deviation. Even when there is no literature studying related interventions, researchers still know that in general, educational interventions rarely raise test scores by more than 50\% of a standard deviation, and extremely rarely raise them by more than one standard deviation. Researchers also know that for binary outcomes, CATEs larger, in absolute value, than 20 or 30 percentage points, are very rare in practice.

\medskip
Under the assumption that $|\tau_s|\leq B$, I derive a closed-form expression of the minimax-linear estimator of $\tau$. It is a weighted sum of the CATEs estimators, with positive weights that sum to less than 1. The most precise estimators receive a weight equal to $p_s$. The least precise estimators receive a weight proportional to one over their variance, and shrunk towards zero. Then, using similar ideas as in \cite{donoho1994statistical}, \cite{armstrong2018}, and \cite{armstrong2021sensitivity}, I outline a simple numerical procedure to approximate CIs for $\tau$ with minimax length.

\medskip
Then, I consider alternative restrictions on the CATEs. First, I assume that $0\leq \tau_s\leq B$ instead of $|\tau_s|\leq B$: on top of being bounded, all CATEs are assumed to be positive. Interestingly, doing so leads to the same minimax-linear estimator of $\tau$ as before. On the other hand, it may lead to a different CI, and I derive, in closed form, confidence lower bounds for $\tau$ with minimax expected excess length.

\medskip
Another alternative restriction I consider is $|\tau_s-\tau|/|\tau|\leq B$. This requires that the distance between $\tau_s$ and $\tau$ cannot be more than $B$ times larger than $|\tau|$, thus restricting CATEs' heterogeneity. $|\tau_s-\tau|/|\tau|\leq B$ implies that CATEs' coefficient of variation (their standard deviation divided by $|\tau|$), a unitless measure of their heterogeneity, is lower than $B$. Deriving the minimax-linear estimator of $\tau$ under the assumption that $|\tau_s-\tau|/|\tau|\leq B$ is considerably more difficult than before, and I am unable to derive a general closed-form. Yet, I can derive a closed-form in the particular case where the largest variance of the CATEs is smaller than $(B+1)$ times the average of their variances. Then, the minimax estimator shrinks all CATEs towards zero uniformly, multiplying them by the same constant, irrespective of their variances.

\medskip
Finally, as minimax estimators can lead to inadmissible tests \cite[][]{lehmann2005testing}, I derive two necessary conditions that minimax-linear estimators and CIs need to satisfy to not lead to inadmissible tests of $\tau=0$, a null of particular interest in treatment-effect estimation. First, they should not assign a weight strictly lower that $p_s$ to all CATE estimators. 
As explained above, if one assumes that $|\tau_s-\tau|/|\tau|\leq B$, the minimax-linear estimator can strictly downweight all CATE estimators. Second, the sum of the weights that a minimax estimator assigns to all CATEs has to be larger than the ratio of its standard error and the standard error of the unbiased estimator. I also show how researchers can estimate the power gain induced by a minimax estimator or CI.

\medskip
I use my results to revisit \cite{behaghel2017ready}, an SRCT, and \cite{connors1996effectiveness}, a matching study. In both cases, assuming that $|\tau_s|\leq B$ for arguably large values of $B$ (50\% of a standard deviation in \cite{behaghel2017ready}, who study an educational intervention, and $0.2$ in \cite{connors1996effectiveness}, whose outcome is binary), I find that the minimax-linear estimator is very close to the unbiased one, but its standard deviation is around 10\% smaller. This gain in precision is not entirely offset by the bias of the minimax-linear estimator: its estimated RMSE is still around 10\% smaller than that of the unbiased estimator, while its worst-case RMSE is around 5-6\% smaller. Minimax CIs are 5 to 7\% smaller than CIs based on the unbiased estimator, and their estimated power to reject the null that $\tau=0$ is 4\% larger in the first application and 1\% larger in the second one (in that application, power is already very high with the unbiased estimator). Results are robust to other choices of $B$ (60\% of a sd in \cite{behaghel2017ready}, 0.3 in \cite{connors1996effectiveness}). Finally, assuming that $|\tau_s-\tau|/|\tau|\leq B_1$ leads, if anything, to slightly smaller precision gains than assumming that $|\tau_s|\leq B_2:=B_1|\tau|$, where letting the second tuning parameter $B_2$ be equal to the first multiplied by $\tau$ ensures that the two assumptions imply comparable restrictions on the CATEs. Overall, minimax estimators and CI lead to a small but non-negligible precision gain. For instance, in an SRCT, achieving a 5\% precision gain by controlling for covariates requires that they account for $10$\% of the outcome's variance. In realistic simulations where I reassign the treatment in the data of \cite{behaghel2017ready}, minimax estimators lead to larger precision and power gains, of 20 and 40\% respectively.

\subsection*{Related literature}

This paper is related to the literature trying to improve the precision of estimators under a conditional independence assumption. A closely related paper is \cite{crump2009}, who assume homoscedasticity and redefine the target parameter as the ATE in the subpopulation whose ATE can be estimated most precisely, which, under some assumptions, are observations with a propensity score in the $[0.1,0.9]$ interval. With respect to that approach, the minimax-linear estimator and CIs I propose do not change the goalpost. This might be an advantage, as the ATE may be an easier-to-motivate target parameter than the ATE in the subsample with a propensity score in the $[0.1,0.9]$ interval. Remarkably, in my second empirical application the minimax-linear estimator is almost as precise as the trimming estimator of \cite{crump2009}, while it does not change the goalpost.

\medskip
An even more closely related paper, posterior to my paper, is \cite{kwon2025estimating}, who derive minimax length CIs for the ATE under the assumption that the variance of the CATEs is bounded: $\sum_{s=1}^Sp_s(\tau_s-\tau)^2\leq B$. With respect to the first restriction on the CATEs I consider ($|\tau_s|\leq B$), applied researchers may find it harder to choose an ex-ante plausible bound for $\sum_{s=1}^Sp_s(\tau_s-\tau)^2$ than for $|\tau_s|$. The reason is simply that until now, researchers have rarely estimated the variance of CATEs, and accordingly they may not have a good ex-ante sense of plausible upper bounds for that parameter.\footnote{This may change now that estimators of $\sum_{s=1}^Sp_s(\tau_s-\tau)^2$ are available and coded in commonly-used statistical software \cite[][]{kline2020leave,de2024estimating}. Researchers could use those tools to estimate $\sum_{s=1}^Sp_s(\tau_s-\tau)^2$ in their data, and calibrate their choice of $B$ to that estimator, but as shown by \cite{armstrong2018optimal}, CIs with a data-driven choice of $B$ will either fail to substantively improve
upon the minimax CIs with $B$ chosen a priori, or fail to maintain coverage over the whole parameter space.} Rather than committing to one value of $B$, one may conduct a sensitivity analysis. One computes CIs for $\tau$ assuming that $\sum_{s=1}^Sp_s(\tau_s-\tau)^2\leq B$ for many values of $B$, until one finds, say, the lowest $B$ such that $0$ belongs to the CI. However, assessing whether this value is plausible for $\sum_{s=1}^Sp_s(\tau_s-\tau)^2$ would again require that researchers have a sense of plausible values for the variance of CATEs, which they currently may not have.


\medskip
This paper is also related to a vast literature in statistics, that has studied minimax-linear and minimax-affine estimators in a bounded-normal-mean model, where realizations of normal variables are used to estimate a linear combination of their bounded means. When I assume that $|\tau_s|\leq B$, the setting I consider can be cast as a bounded-mean model.\footnote{I do not assume that CATEs estimators are normally distributed,  but as noted by \cite{armstrong2018finite}, this distributional assumption is not of essence to derive the minimax-linear estimator.}
\cite{donoho1994statistical} shows that, in a model nesting the bounded-normal-mean one, the risk of the minimax-affine estimator cannot be more than 25\% larger than that of the minimax estimator, thus motivating the study of minimax-affine and minimax-linear estimators. The closed-form expression of the minimax-linear estimator I derive assuming that $|\tau_s|\leq B$ has, to my knowledge, not been derived earlier in that literature. 

\medskip
Finally, my paper is also related to a growing econometrics literature that has applied the set-up in \cite{donoho1994statistical} to treatment-effect estimation, including: \cite{armstrong2018finite}, who study ATE estimation under uncounfoundedness when the mean outcome conditional on the covariates is Lipschitz with a bounded Lipschitz constant; \cite{armstrong2021sensitivity}, who study  sensitivity analysis in locally misspecified GMM models; \cite{rambachan2019honest}, who study difference-in-differences estimation with bounded departures from parallel trends; \cite{armstrong2018optimal}, \cite{imbens2019optimized}, and \cite{noack2019bias}, who study estimation in regression discontinuity designs with bounded second derivatives of mean potential outcomes conditional on the running variable.

\paragraph{Organization of the paper.} Section 2 introduces the setup. Section 3 presents the minimax-linear estimator under the assumption that $|\tau_s|\leq B$. Section 4 presents minimax-linear estimators under alternative restrictions. Section 5 presents minimax confidence intervals. Section 6 discusses the power of minimax-linear estimators. Section 7 presents the empirical applications. Section 8 presents some simulations. Proofs are in the appendix.

\section{Setup and examples}

\begin{definition} (ATE and CATEs)\label{def:setup}
One is interested in estimating an average treatment effect $\tau$, equal to a weighted average of $S$ conditional average treatment effects $(\tau_s)_{1\leq s\leq S}$, with weights $(p_s)_{1\leq s\leq S}$ that are known, positive, and sum to 1:
\begin{equation}\label{eq:decomposition}
    \tau =  \sum_{s=1}^Sp_s\tau_s.
\end{equation}
\end{definition}
\begin{hyp}\label{hyp:CATE_hat}
One observes random variables $(\widehat{\tau}_s)_{1\leq s\leq S}$ such that:
    \begin{enumerate}
        \item \label{hyp:CATE_hat_p1} $E\left(\widehat{\tau}_s\right)=\tau_s$ for all $s$.
        \item \label{hyp:CATE_hat_p2} $cov\left(\widehat{\tau}_s,\widehat{\tau}_{s'}\right)=0$ for all $s'\ne s$.
        \item \label{hyp:CATE_hat_p3} $V_s:=V\left(\widehat{\tau}_s\right)<+\infty$.
        \end{enumerate}
\end{hyp}
Points \ref{hyp:CATE_hat_p1} to \ref{hyp:CATE_hat_p3} require that the CATEs can be unbiasedly estimated, with estimators that are uncorrelated across $s$ and have a finite variance. Those conditions are satisfied in a number of research designs, as the two examples below show.

\begin{example}\label{ex:SRCTs}
\textbf{: SRCTs.} Consider an SRCT with $S$ strata indexed by $s$. Let $\tau_s$ be the CATE in stratum $s$, let $p_s$ be the share of the population stratum $s$ accounts for, and let $\tau$ be the ATE. Let $\widehat{\tau}_s$ be the difference between the average outcome of treated and control units in stratum $s$. Let $n_s$ be the number of units in stratum $s$, let $n_{1,s}\in \{1,...,n_s-1\}$ be the number of treated units, let $n_{0,s}=n_s-n_{1,s}$ be the number of control units, let $\textbf{D}_s$  be a vector stacking the treatment indicators of all units in stratum $s$, and let $\textbf{Y}_s$  be a vector stacking their potential outcomes. Assume that:
\begin{enumerate}
\item for all $s$, for every $(d_1,...,d_{n_s})\in \{0,1\}^{n_s}$ such that $d_{1}+...+d_{n_s}=n_{1,s}$, $$P(\textbf{D}_s=(d_1,...,d_{n_s})|\textbf{Y}_s)=\frac{1}{{n_s \choose n_{1,s}}}.$$
\item The random vectors $(\textbf{D}_s,\textbf{Y}_s)$ are mutually independent across $s$.
\end{enumerate}
Points 1 and 2 above are standard conditions that hold by design in SRCTs. They imply that Points 1 and 2 of Assumption \ref{hyp:CATE_hat} hold \cite[see e.g.][]{imbens2015}. If one further assumes that units in stratum $s$ are randomly drawn from a superpopulation, then
\begin{equation}\label{eq:Vs_example1}
V_s=\frac{\sigma^2_{0,s}}{n_{0,s}}+\frac{\sigma^2_{1,s}}{n_{1,s}},
\end{equation}
where $\sigma^2_{0,s}$ and $\sigma^2_{1,s}$ respectively denote the variances of the outcomes without and with treatment in the superpopulation \cite[see e.g.][]{imbens2015}. Therefore, Point \ref{hyp:CATE_hat_p3} of Assumption \ref{hyp:CATE_hat} holds if $\sigma^2_{0,s}<+\infty$ and $\sigma^2_{1,s}<+\infty$ for all $s$.
\end{example}

\begin{example}\label{ex:matching}
\textbf{: estimation under conditional independence, using an augmented inverse propensity weighting (AIPW) estimator.}
Let $s$ index independent and identically distributed (iid) units drawn from a superpopulation. Let $Y_s(0)$ and $Y_s(1)$ denote the untreated and treated outcomes of $s$, let $D_s$ and $Y_s$ denote their treatment status and observed outcome, and let $X_s$ denote a vector of covariates. Assume that
\begin{equation}\label{eq:ignorable_treatment}
(Y_{s}(0),Y_{s}(1))\indep D_s|X_s:
\end{equation}
treatment is ignorable conditional on the covariates.
Let $\tau_s=E(Y_s(1)-Y_s(0)|X_s)$, and let $p_s=1/S$. Then, $\tau=1/S\sum_{s=1}^S E(Y_s(1)-Y_s(0)|X_s)$ is the sample average treatment effect, a parameter also considered by \cite{crump2009}. Let $e(X_s)=P(D_{s}=1|X_s)$ denote the propensity-score, and for all $d\in \{0,1\}$, let $\mu_d(X_s)=E(Y_s|D_s=d,X_s)$. Then, let
\begin{equation}\label{eq:tau_s_DR}
\widehat{\tau}_s=\mu_1(X_s)-\mu_0(X_s)+D_s\frac{Y_s-\mu_1(X_s)}{e(X_s)}-(1-D_s)\frac{Y_s-\mu_1(X_s)}{1-e(X_s)}.
\end{equation}
$1/S\sum_{s=1}^S \widehat{\tau}_s$ is the oracle version of the AIPW estimator of \cite{robins1994estimation}.
Under \eqref{eq:ignorable_treatment},
$$E(\widehat{\tau}_s|X_s)=\tau_s,$$
so Point \ref{hyp:CATE_hat_p1} of Assumption \ref{hyp:CATE_hat} holds conditional on $X_s$.
Point \ref{hyp:CATE_hat_p2} mechanically holds as units are assumed to be iid.
If for all $d\in \{0,1\}$, $V(Y_s(d)|X_s)=\sigma_d^2(X_s)<+\infty$, then
\begin{equation}\label{eq:Vs_example2}
V\left(\widehat{\tau}_s|X_s\right)=\frac{\sigma_0^2(X_s)}{1-e(X_s)}+\frac{\sigma_1^2(X_s)}{e(X_s)},
\end{equation}
so Point \ref{hyp:CATE_hat_p3} of Assumption \ref{hyp:CATE_hat} also holds conditional on $X_s$, with $V_s=\frac{\sigma_0^2(X_s)}{1-e(X_s)}+\frac{\sigma_1^2(X_s)}{e(X_s)}$.
\end{example}



\section{Minimax-linear estimator of $\tau$ with bounded CATEs}

In this section, I assume that all CATEs are smaller in absolute value than a constant $B$.
\begin{hyp}\label{hyp:CATE_hat_p4} There is a known $B\in \mathbb{R}_+\setminus\{0\}$ such that $|\tau_s|\leq B$ for all $s$.
\end{hyp}
For any $1\times S$ deterministic vector $\bm{w}=(w_1,...,w_S)$, let
\begin{equation}\label{eq:lin_comb_CATE}
\widehat{\tau}(\bm{w})=\sum_{s=1}^Sw_s\widehat{\tau}_s.
\end{equation}
$\widehat{\tau}(\bm{w})$ is a linear combination of the estimators $\widehat{\tau}_s$.
Lemma \ref{lem_worst-case-MSE} gives its worst-case MSE.
\begin{lem}\label{lem_worst-case-MSE} (Worst-case MSE of $\widehat{\tau}(\bm{w})$)

If Assumptions \ref{hyp:CATE_hat} and \ref{hyp:CATE_hat_p4} hold,
$$E\left(\left(\widehat{\tau}(\bm{w})-\tau\right)^2\right)\leq \overline{\text{MSE}}(\bm{w}):= \sum_{s=1}^Sw_s^2V_s+B^2\left(\sum_{s=1}^S|w_s-p_s|\right)^2.$$
The upper bound in the previous display is sharp.
\end{lem}
Without loss of generality, assume that
$$p_1V_1\leq p_2V_2 \leq ... \leq p_SV_S.$$
Let $$\underline{s}=\min \left\{s \in \{1,...,S\}: \frac{1}{\frac{1}{B^2}+\sum_{s'=s}^S\frac{1}{V_{s'}}}\sum_{s'=s}^Sp_{s'}< p_s V_s\right\}.$$ $\underline{s}$ is well defined, because $$\frac{1}{\frac{1}{B^2}+\frac{1}{V_{S}}}p_{S}< p_S V_S.$$

Let
\begin{align}
&w^M_{B,s}=p_s \text{ for all } s<\underline{s} \nonumber\\
&w^M_{B,s}=\frac{1}{V_s}\frac{1}{\frac{1}{B^2}+\sum_{s'=\underline{s}}^S\frac{1}{V_{s'}}}\sum_{s'=\underline{s}}^Sp_{s'} \text{ for all } s\geq \underline{s}. \label{eq:w_opt2}
\end{align}

\begin{thm}\label{thm_minimax}
If Assumption \ref{hyp:CATE_hat} holds, $\bm{w}^M_B=\argmin_{\bm{w}\in \mathbb{R}^S} \overline{\text{MSE}}(\bm{w}).$
\end{thm}
Theorem \ref{thm_minimax} shows that $\widehat{\tau}(\bm{w}^M_B)$ is the minimax-linear estimator of $\tau$ under Assumption \ref{hyp:CATE_hat}. That estimator is a weighted sum of the $\widehat{\tau}_s$s, with positive weights that sum to less than 1. For a precisely estimated $\widehat{\tau}_s$ (one with a low value of $p_sV_s$), the optimal weight is just $p_s$. On the other hand, for an imprecisely estimated $\widehat{\tau}_s$, the optimal weight is proportional to one over its variance. The sum of the weights on imprecise CATEs is equal to
$$\frac{\sum_{s=\underline{s}}^S\frac{1}{V_{s}}}{\frac{1}{B^2}+\sum_{s=\underline{s}}^S\frac{1}{V_{s}}}\sum_{s=\underline{s}}^Sp_{s}<\sum_{s=\underline{s}}^Sp_{s},$$ 
so the extent to which those estimators are downweighted depends on $B$. 
$\forall s~\underset{B\rightarrow +\infty}{\lim} w^M_{B,s}=p_s:$ the minimax estimator converges to the unbiased one when CATEs become unbounded.

\paragraph{Feasible estimator.}
In general, $\widehat{\tau}(\bm{w}^M_B)$ is infeasible, as it depends on unknown quantities. A feasible estimator can be obtained, by replacing those quantities by estimators. First, the optimal weights $w^M_{B,s}$ depend on $(V_s)_{1\leq s \leq S}$. In Example 1, if the SRCT has at least two treated and two control units per stratum, to estimate $V_s$ one can replace $\sigma^2_{0,s}$ and $\sigma^2_{1,s}$ in \eqref{eq:Vs_example1} by the variances of the outcome in the control and treatment groups, respectively. Similarly, in Example 2, to estimate $V_s$ one can replace $\sigma_0^2(X_s)$, $\sigma_1^2(X_s)$, and $e(X_s)$ in \eqref{eq:Vs_example2} by some estimators. Second, in Example 2 one also needs to estimate $\widehat{\tau}_s$, something that can be done replacing $\mu_0(X_s)$, $\mu_1(X_s)$, and $e(X_s)$ by estimators in \eqref{eq:tau_s_DR}.



\subsection{In an SRCT, the minimax estimator is feasible under homoscedasticity}

In this subsection, we assume we are in an SRCT with at least two treated and two control units per stratum.
Let $v_{s}=1/n_{0,s}+1/n_{1,s}$.
Without loss of generality, assume that
$$p_1v_1\leq p_2v_2 \leq ... \leq p_Sv_S.$$
Let $\underline{s}^{\text{hom}}=\min \left\{s \in \{1,...,S\}: \frac{1}{\frac{1}{B^2}+\sum_{s'=s}^S\frac{1}{v_{s'}}}\sum_{s'=s}^Sp_{s'}< p_s v_s\right\}$, and let
\begin{align*}
&w^{\text{hom}}_{B,s}=p_s \text{ for all } s<\underline{s}^{\text{hom}} \nonumber\\
&w^{\text{hom}}_{B,s}=\frac{1}{v_s}\frac{1}{\frac{1}{B^2}+\sum_{s'=\underline{s}^{\text{hom}}}^S\frac{1}{v_{s'}}}\sum_{s'=\underline{s}^{\text{hom}}}^Sp_{s'} \text{ for all } s\geq \underline{s}^{\text{hom}}.
\end{align*}
It follows from the definition of $\left(w^M_{B,s}\right)_{s \in \{1,...,S\}}$ that under the following homoscedasticity condition:
\begin{equation}\label{eq:homoscedasticity}
\sigma^2_{0,s}=\sigma^2_{1,s}=\sigma^2,
\end{equation}
$\bm{w}^M_{\sigma B}=\bm{w}^{\text{hom}}_{B}$. Then, if the researcher assumes that CATEs are all lower than $B$ standard deviation of the outcome, the minimax-linear estimator is feasible given $B$, as its weights only depend on known quantities.

\paragraph{Properties of $\widehat{\tau}\left(\bm{w}^{\text{hom}}_{B}\right)$ with heteroscedasticity.}
Of course, the homoscedasticity assumption underlying $\widehat{\tau}\left(\bm{w}^{\text{hom}}_{B}\right)$ is strong. Yet, I now give sufficient conditions under which $\widehat{\tau}\left(\bm{w}^{\text{hom}}_{B}\right)$ has lower worst-case MSE than $\widehat{\tau}\left(\bm{p}\right)$, even if the outcome is heteroscedastic. Let $v_{0,s}=1/n_{0,s}$ and $v_{1,s}=1/n_{1,s}$.
\begin{cor}\label{cor_heteroscedasticite}
\begin{enumerate}
\item If Assumption \ref{hyp:CATE_hat} holds, and if for all $s\in\{1,...,S\}$ $\sigma^2=\sigma^2_{0,s}\leq \sigma^2_{1,s}$, then
the worst-case MSE of $\widehat{\tau}\left(\bm{w}^{\text{hom}}_{B}\right)$ is lower than that of $\widehat{\tau}\left(\bm{p}\right)$.
\item  If Assumption \ref{hyp:CATE_hat} holds, and if for all $s\in\{1,...,S\}$ $\sigma^2_{0,s}=\sigma^2$ and $\sigma^2_{1,s}=h\sigma^2$, then if
    $$h\geq \frac{B^2\left(\sum_{s=1}^S\left(p_s-w^{\text{hom}}_{B,s}\right)\right)^2-\sum_{s=1}^S\left((p_s)^2-\left(w^{\text{hom}}_{B,s}\right)^2\right)v_{0,s}}{\sum_{s=1}^S\left((p_s)^2-\left(w^{\text{hom}}_{B,s}\right)^2\right)v_{1,s}},$$
    the worst-case MSE of $\widehat{\tau}\left(\bm{w}^{\text{hom}}_{B}\right)$ is lower than that of $\widehat{\tau}\left(\bm{p}\right)$.
\end{enumerate}
\end{cor}
Point 1 of Corollary \ref{cor_heteroscedasticite} assumes that the untreated outcome's variance does not vary across strata. This for instance holds when in each stratum, researchers standardize their outcome by its standard deviation among the stratum's control group. Point 1 of Corollary \ref{cor_heteroscedasticite} further assumes  that in each stratum, the variance of the treated outcome is larger than that of the untreated one. Under these two conditions, the worst-case MSE of $\widehat{\tau}\left(\bm{w}^{\text{hom}}_{B}\right)$ is lower than that of $\widehat{\tau}\left(\bm{p}\right)$. Intuitively, $\widehat{\tau}\left(\bm{w}^{\text{hom}}_{B}\right)$ underestimates the variances of all the CATE estimators, which leads it to not shrink those estimators enough, but it still dominates the
unbiased estimator that does not do any shrinkage. Point 2 of Corollary \ref{cor_heteroscedasticite} assumes that the variances of the untreated and treated outcomes do not vary across strata. Under these conditions, the worst-case MSE of $\widehat{\tau}\left(\bm{w}^{\text{hom}}_{B}\right)$ is lower than that of $\widehat{\tau}\left(\bm{p}\right)$, provided that the ratio of the treated and untreated outcomes' variances is greater than a bound which only depends on the design and can be readily computed. In the first SRCT I revisit in Section \ref{sec:application}, this lower bound is negative so the worst-case MSE of $\widehat{\tau}\left(\bm{w}^M_B\right)$ is guaranteed to be lower than that of $\widehat{\tau}\left(\bm{p}\right)$ if either the assumptions of Point 1 or Point 2 of Corollary \ref{cor_heteroscedasticite} hold.\footnote{In Corollary \ref{cor_heteroscedasticite}, $\sigma$ represents the standard deviation of the untreated outcome, and $\widehat{\tau}\left(\bm{w}^{\text{hom}}_{B}\right)$ therefore assumes that the CATEs are bounded by $B\%$ of the untreated outcome's variance. If one uses instead the standard deviation of the treated outcome  as the numeraire, the conclusions of Corollary \ref{cor_heteroscedasticite} revert. For instance, $\widehat{\tau}\left(\bm{w}^{\text{hom}}_{B}\right)$'s worst-case MSE is always lower than $\widehat{\tau}\left(\bm{p}\right)$'s if the untreated outcome's variance is larger than that of the treated outcome. Using the standard deviation of the untreated outcome  as the numeraire follows the common practice in applied work of standardizing the outcome by its standard deviation in the control group.}

\paragraph{Estimating the variance of $\widehat{\tau}\left(\bm{w}^{\text{hom}}_{B}\right)$.}
As given $B$, the weights $\bm{w}^{\text{hom}}_{B}$ are not stochastic, it is easy to show that under Assumption \ref{hyp:CATE_hat},
\begin{equation}\label{eq:conservative_variance_estimator}
V\left(\widehat{\tau}\left(\bm{w}^{\text{hom}}_{B}\right)\right)=\sum_{s=1}^S\left(w^{\text{hom}}_{B,s}\right)^2\left(\sigma^2_{0,s}/n_{0,s}+\sigma^2_{1,s}/n_{1,s}\right).
\end{equation}
Importantly, \eqref{eq:conservative_variance_estimator} holds even if the outcome is heteroscedastic. The right hand side of the previous display can easily be estimated.

\section{Alternative restrictions on the CATEs}

\subsection{Assuming that CATEs are bounded and are all of the same, known sign}

In this section, we replace Assumption \ref{hyp:CATE_hat_p4} by:
\begin{hyp}\label{hyp:CATE_hat_sign}
There is a known $B\in \mathbb{R}_+\setminus\{0\}$ such that $0\leq \tau_s \leq B$ for all $s$.
\end{hyp}
On top of bounding the size of the CATEs, like Assumption \ref{hyp:CATE_hat_p4}, Assumption \ref{hyp:CATE_hat_sign} also assumes they are all positive (what follows still applies if we instead assume they are all negative).

\medskip
Let $\bm{w}=(w_1,...,w_S)$ be an arbitrary vector in $\mathbb{R}^S$. For any real number $x$, let $x_+=\max(x,0)$ and let $x_-=\min(x,0)$.
\begin{lem}\label{lem_worst-case-MSE-sign} (Worst-case MSE of $\widehat{\tau}(\bm{w})$ under Assumption \ref{hyp:CATE_hat_sign})
If Assumptions \ref{hyp:CATE_hat} and \ref{hyp:CATE_hat_sign} hold,
\begin{align*}
E\left(\left(\widehat{\tau}(\bm{w})-\tau\right)^2\right)\leq &\overline{\text{MSE}}^{\text{MS}}(\bm{w}):= \sum_{s=1}^Sw_{s}^2V_{s}+B^2\max\left[\left(\sum_{s=1}^{S}(w_s-p_s)_+\right)^2,\left(\sum_{s=1}^{S}(w_s-p_s)_-\right)^2\right].
\end{align*}
The upper bound is sharp.
\end{lem}
The worst-case MSEs of $\widehat{\tau}(\bm{w})$ under Assumptions \ref{hyp:CATE_hat_p4} and \ref{hyp:CATE_hat_sign} differ. However, this difference is inconsequential. In the proof of Theorem \ref{thm_minimax}, I show that under Assumption  \ref{hyp:CATE_hat_p4}, the weights of the minimax-linear estimator is the minimizer of $\overline{\text{MSE}}^{\text{MS}}(\bm{w})$ across all $\bm{w}$ such that $w_s\leq p_s$ for all $s\in \{1,...,S\}$. Similarly, assume that $\bm{w}^{\text{MS}}$, the minimizer of $\overline{\text{MSE}}^{\text{MS}}(\bm{w})$, has at least one coordinate that is strictly larger than the corresponding coordinate of $(p_1,...,p_S)$. Without loss of generality, assume that $w^{\text{MS}}_{1}>p_1$. Then,
$\overline{\text{MSE}}^{\text{MS}}(\bm{w}^{MS})>\overline{\text{MSE}}^{\text{MS}}(p_1,w^{MS}_{2},...,w^{MS}_{S})$, a contradiction. Therefore, $\bm{w}^{MS}_s\leq p_s$ for all $s$. Accordingly, the weights of the minimax-linear estimator under Assumption \ref{hyp:CATE_hat_sign} is the minimizer of $\overline{\text{MSE}}^{\text{MS}}(\bm{w})$, across all $\bm{w}$ such that $w_s\leq p_s$ for all $s\in \{1,...,S\}$. But if $w_s\leq p_s$ for all $s\in \{1,...,S\}$, $\overline{\text{MSE}}^{\text{MS}}(\bm{w})=\overline{\text{MSE}}(\bm{w})$.  Therefore, the minimax-linear estimators under Assumptions \ref{hyp:CATE_hat_p4} and \ref{hyp:CATE_hat_sign} are equal.

\subsection{Bounding CATEs' heterogeneity}

Throughout this section, I make the following assumption on the design.
\begin{hyp}\label{hyp:design}
$S$ is even, $p_s=1/S$, and $V_{s}<V_{s+1}$ for all $s\leq S-1$.
\end{hyp}
Assuming that $S$ is even simplifies the analysis and is without great loss of generality. $p_s=1/S$ implies that results below apply to Example \ref{ex:matching} but not to Example \ref{ex:SRCTs}. $V_{s}<V_{s+1}$ for instance holds in Example \ref{ex:matching} if $\sigma^2_d(X_s)=\sigma^2$ for all $d$ and $X_s$, and $e(X_s)\ne e(X_{s'})$ for all $s\ne s'$ (this second condition for instance holds if the propensity score follows, say, a logit model, and $X_s$ is a scalar continuously distributed variable with a non-zero coefficient).

\medskip
Then, I replace Assumption \ref{hyp:CATE_hat_p4} by:
\begin{hyp}\label{hyp:CATE_hat_het}
$\tau\ne 0$, and there is a known $B\in \mathbb{R}_+$ such that
$\left|\frac{\tau_s-\tau}{\tau}\right| \leq B$ $\forall s$.
\end{hyp}
Assumption \ref{hyp:CATE_hat_p4} requires that the distance between $\tau_s$ and $\tau$ cannot be more than $B$ times larger than $|\tau|$, thus restricting CATEs' heterogeneity. It implies that $B$ is larger than CATEs' coefficient of variation:
\begin{equation}\label{eq:bound_variance}
\frac{\sqrt{1/S\sum_{s=1}^S\left(\tau_s-\tau\right)^2}}{|\tau|}\leq B.
\end{equation}
With $B=0$, Assumption \ref{hyp:CATE_hat_p4} requires that all CATEs are equal. With $B\leq 1$, Assumption \ref{hyp:CATE_hat_p4} requires that all CATEs are of the same sign.

\medskip
Let $\bm{w}=(w_1,...,w_S)$ be an arbitrary vector in $\mathbb{R}^S_+$, and let $(s)$ denote a permutation of $\{1,...,S\}$ such that $s\mapsto w_{(s)}$ is decreasing. $(s)$ is a function of $\bm{w}$, but for now we leave that dependence implicit. That permutation may not be unique, but results below hold for any permutation such that $s\mapsto w_{\sigma(s)}$ is decreasing.
\begin{lem}\label{lem_worst-case-MSE_boundhet} (Worst-case MSE of $\widehat{\tau}(\bm{w})$ under Assumption \ref{hyp:CATE_hat_het})
Assume that Assumptions \ref{hyp:CATE_hat}, \ref{hyp:design}, and \ref{hyp:CATE_hat_het} hold. Then,
\begin{align*}
&E\left(\left(\widehat{\tau}(\bm{w})-\tau\right)^2\right)\\
\leq &\sum_{s=1}^Sw_{s}^2V_{s}+\tau^2\max\left[\left((B+1)\sum_{s=1}^{S/2}(w_{(s)}-1/S)-(B-1)\sum_{s=S/2+1}^{S}(w_{(s)}-1/S)\right)^2,\right.\\
&\left.\left((B+1)\sum_{s=S/2+1}^{S}(w_{(s)}-1/S)-(B-1)\sum_{s=1}^{S/2}(w_{(s)}-1/S)\right)^2\right].
\end{align*}
The upper bound is sharp.
\end{lem}
The proof of Lemma \ref{lem_worst-case-MSE_boundhet} amounts to showing that for any $\bm{w}$, the worst-case MSE is reached letting either
$\tau^+_{(s)}=(B+1)\tau1\{s\leq S/2\}-(B-1)\tau1\{s> S/2\}$ or $\tau^-_{(s)}=(B+1)\tau1\{s> S/2\}-(B-1)\tau1\{s\leq S/2\}$. In the first scenario, the CATEs of the half of the sample for which $w_{s}$ is the largest are all equal to $(B+1)\tau$, while the CATEs of the half of the sample for which $w_{s}$ is the lowest are all equal to $-(B-1)\tau$. In the second scenario, the CATEs of the half of the sample for which $w_{s}$ is the lowest are all equal to $(B+1)\tau$, while the CATEs of the half of the sample for which $w_{s}$ is the largest are all equal to $-(B-1)\tau$.

\medskip
Let
\begin{align*}
V(\bm{w})=& \sum_{s=1}^Sw_{s}^2V_s,\\
\overline{SQB}^H(\bm{w})=&\tau^2\max\left[\left((B+1)\sum_{s=1}^{S/2}(w_{(s)}-1/S)-(B-1)\sum_{s=S/2+1}^{S}(w_{(s)}-1/S)\right)^2,\right.\\
&\left.\left((B+1)\sum_{s=S/2+1}^{S}(w_{(s)}-1/S)-(B-1)\sum_{s=1}^{S/2}(w_{(s)}-1/S)\right)^2\right],\\
\overline{\text{MSE}}^H(\bm{w})=& V(\bm{w})+\overline{SQB}^H(\bm{w}).
\end{align*}
Let\footnote{In this section, the fact that the optimal weights may depend on $B$ is left implicit.}
$$\bm{w}^H=\argmin_{\bm{w}\in \mathbb{R}^S_+} \overline{\text{MSE}}^H(\bm{w}).$$
Under Assumptions \ref{hyp:CATE_hat}, \ref{hyp:design}, and \ref{hyp:CATE_hat_het}, $\widehat{\tau}(\bm{w}^H)$ is the minimax estimator of $\tau$, across all linear combinations of $\widehat{\tau}_s$ with positive weights. Restricting attention to linear combinations of $\widehat{\tau}_s$ with positive weights may be appealing: linear combinations with negative weights could suffer from sign reversal phenomena, where, say, $E\left(\widehat{\tau}(\bm{w})\right)<0$ even if $\tau_s>0$ for all $s$.

\medskip
We now state two lemmas that will allow us to show that $\bm{w}^H$ is the minimizer of a convex quadratic function subject to linear equality constraints. 
Henceforth, we let $[s]$ denote a permutation such that $s\mapsto w^H_{[s]}$ is decreasing.
\begin{lem}\label{lem_average_opt_weights_lowerthanone}
\begin{enumerate}
\item $\sum_{s=1}^{S}w^H_{[s]}\leq 1.$
\item For any $\bm{w}\in \mathbb{R}^S$, if $\sum_{s=1}^{S}w_{(s)}\leq 1$ then
\begin{align*}
&\left|(B+1)\sum_{s=1}^{S/2}(w_{(s)}-1/S)-(B-1)\sum_{s=S/2+1}^{S}(w_{(s)}-1/S)\right|\\
\leq & \left|(B+1)\sum_{s=S/2+1}^{S}(w_{(s)}-1/S)-(B-1)\sum_{s=1}^{S/2}(w_{(s)}-1/S)\right|.
\end{align*}
\end{enumerate}
\end{lem}
\begin{lem}\label{lem_permutation}
Assume that $V_{s}<V_{s+1}$ for all $s\leq S-1$. Then, $w^H_{s}\geq w^H_{s+1}$ for all $s\leq S-1$.
\end{lem}
Heuristically, Lemma \ref{lem_average_opt_weights_lowerthanone} implies that $\overline{SQB}^H(\bm{w}^H)$
is actually just equal to
\begin{align*}
&\left((B+1)\sum_{s=S/2+1}^{S}(w^H_{[s]}-1/S)-(B-1)\sum_{s=1}^{S/2}(w^H_{[s]}-1/S)\right)^2,
\end{align*}
thus allowing us to get rid of the non-differentiable $\max$ operator in $\overline{\text{MSE}}^H(\bm{w})$, while Lemma \ref{lem_permutation} implies that $s\mapsto w^H_{s}$ is decreasing, so that
\begin{align*}
&\left((B+1)\sum_{s=S/2+1}^{S}(w^H_{[s]}-1/S)-(B-1)\sum_{s=1}^{S/2}(w^H_{[s]}-1/S)\right)^2\\
=&\left((B+1)\sum_{s=S/2+1}^{S}(w^H_{s}-1/S)-(B-1)\sum_{s=1}^{S/2}(w^H_{s}-1/S)\right)^2\\
=&\left((B-1)\sum_{s=1}^{S/2}w^H_{s}-(B+1)\sum_{s=S/2+1}^{S}w^H_{s}+1\right)^2.
\end{align*}
Then, let
\begin{align*}
\overline{\text{MSE}}^{H,d}(\bm{w})=&V(\bm{w})+\tau^2\left((B-1)\sum_{s=1}^{S/2}w_{s}-(B+1)\sum_{s=S/2+1}^{S}w_{s}+1\right)^2,
\end{align*}
and let
$$\mathcal{R}:=\left\{\bm{w}\in \mathbb{R}^S:\sum_{s=1}^{S}w_{s}-1\leq 0,~\forall
s \in \{1,...,S-1\}:w_{s+1}-w_{s}\leq 0,~-w_S\leq 0\right\}.$$
\begin{thm}\label{thm_minimax2}
Assume that Assumptions \ref{hyp:CATE_hat}, \ref{hyp:design}, and \ref{hyp:CATE_hat_het} hold. Then:
\begin{enumerate}
\item $\bm{w}^H=\argmin_{\bm{w}\in \mathcal{R}} \overline{\text{MSE}}^{H,d}(\bm{w})$.
\item If $B\geq 1$, $w^H_{1}=w^H_{s_0}$, where $s_0=\lfloor SB/(B+1) \rfloor+1$.
\item If $B\geq 1$ and $\frac{V_S}{1/S\sum_{s'=1}^{S}V_{s'}}\leq B+1$,
    $$w^H_{1}=w^H_{S}=\frac{\tau^2}{\frac{1}{S^2}\sum_{s=1}^{S}V_{s}+\tau^2}1/S.$$
\end{enumerate}
\end{thm}
As explained above, Point 1 of Theorem \ref{thm_minimax2} readily follows from Lemmas \ref{lem_average_opt_weights_lowerthanone} and \ref{lem_permutation}. This result implies that $\bm{w}^H$ can be approximated in polynomial time by standard convex quadratic programming algorithms. Point 2 shows that if $B\geq 1$, then the first $s_0$ coordinates of $\bm{w}^H$ are equal to each other, where $s_0=\lfloor S(1-1/(B+1)) \rfloor+1:$ essentially, only the $1/(B+1)$\% of estimators with the largest variances can receive a strictly lower weight than the other estimators. Then, Point 3 shows that if $B\geq 1$ and if the largest variance is not larger than $(B+1)$ times the average variance, all coordinates of $\bm{w}^H$ are equal. Then, perhaps surprisingly, $\widehat{\tau}(\bm{w}^H)$ shrinks uniformly all estimators $\widehat{\tau}_s$ by the same constant, irrespective of their variance $V_{s}$. As explained in Section \ref{sec:power} below, this implies that the CI for $\tau$ based on $\widehat{\tau}(\bm{w}^H)$ contains zero more often than the CI based on $\widehat{\tau}(\bm{p})$, despite the fact that the former CI might be shorter. Then, using $\widehat{\tau}(\bm{w}^H)$ instead of $\widehat{\tau}(\bm{p})$ will reduce the power to reject the null hypothesis of no treatment effect, thus making $\widehat{\tau}(\bm{w}^H)$ an unappealing alternative to $\widehat{\tau}(\bm{p})$. 

\section{Minimax confidence intervals and confidence bounds}

Throughout this section and the next, I make the following assumption:
\begin{hyp}\label{hyp:CATE_hat_normal}
$\widehat{\tau}(\bm{w})$ follows a normal distribution.
\end{hyp}
Then,
\begin{equation*}
\widehat{\tau}(\bm{w})-\tau \sim \mathcal{N}\left(B(\bm{w}),\sigma^2(\bm{w})\right).
\end{equation*}

\subsection{Assuming that CATEs are bounded}

Under Assumption \ref{hyp:CATE_hat_p4},
\begin{equation}\label{eq:boundbias_1}
\left| B(\bm{w})\right| \leq \overline{|B|}(\bm{w}):=B\sum_{s=1}^S|w_s-p_s|.
\end{equation}
Then, to form a confidence interval (CI) for $\tau$, one can follow \cite{donoho1994statistical,armstrong2018,armstrong2021sensitivity}, and use the fixed-length CI, centered around $\widehat{\tau}(\bm{w})$, with minimax length. As shown by those papers, for a level $1-\alpha$, this minimax CI is equal to $$\widehat{\tau}(\bm{w}^{\text{CI}})+/-Q_{1-\alpha}\left(\overline{|B|}(\bm{w}^{\text{CI}}),\sigma(\bm{w}^{\text{CI}})\right),$$
where $Q_{1-\alpha}(\mu,\sigma)$ denotes the quantile of order $1-\alpha$ of $|X|$ where $X \sim \mathcal{N}\left(\mu,\sigma^2\right)$,
and where $\bm{w}^{\text{CI}}$ is the minimizer of $Q_{1-\alpha}\left(\overline{|B|}(\bm{w}),\sigma(\bm{w})\right)$ across all $\bm{w}\in \mathbb{R}^S$.

\medskip
This minimization problem can be solved using similar ideas as those proposed by \cite{armstrong2018} and \cite{armstrong2021sensitivity}. First, note that $0\leq w^{\text{CI}}_s\leq p_s$.\footnote{\label{footnote_boundsoptimalweights} First, assume that, say, $w^{\text{CI}}_1> p_1$. Let $\tilde{w}_s=p_11\{s=1\}+w^{\text{CI}}_s 1\{s>1\}$. $\overline{|B|}(\bm{\tilde{w}})<\overline{|B|}(\bm{w}^{\text{CI}})$ and $\sigma(\bm{\tilde{w}})<\sigma(\bm{w}^{\text{CI}})$, so $Q_{1-\alpha}\left(\overline{|B|}(\bm{\tilde{w}}),\sigma(\bm{\tilde{w}})\right)<Q_{1-\alpha}\left(\overline{|B|}(\bm{\bm{w}^{\text{CI}}}),\sigma(\bm{\bm{w}^{\text{CI}}})\right),$
a contradiction. Second, assume that, say, $w^{\text{CI}}_1<0$. Let $\tilde{w}_s=w^{\text{CI}}_s 1\{s>1\}$. $\overline{|B|}(\bm{\tilde{w}})<\overline{|B|}(\bm{w}^{\text{CI}})$ and $\sigma(\bm{\tilde{w}})<\sigma(\bm{w}^{\text{CI}})$, so $Q_{1-\alpha}\left(\overline{|B|}(\bm{\tilde{w}}),\sigma(\bm{\tilde{w}})\right)<Q_{1-\alpha}\left(\overline{|B|}(\bm{\bm{w}^{\text{CI}}}),\sigma(\bm{\bm{w}^{\text{CI}}})\right),$
a contradiction.} Therefore, $\overline{|B|}(\bm{w}):=B\sum_{s=1}^S(p_s-w_s)\in [0,B]$. Then, for any $M:0<M<B$, let
$\bm{w}^{\text{CI}}_M$ be the minimizer of $\sigma^2(\bm{w})$ subject to $B\sum_{s=1}^S(p_s-w_s)\leq M$ and $w_s\leq p_s$.
For any $k\in \{1,...,S\}$, let $$\lambda(k)=\frac{\sum_{s=k}^Sp_s-M/B}{B\sum_{s=k}^S1/V_s},$$
and let
$$\mathcal{S}_{\text{ci}}=\{s:\lambda(s)>0,~\lambda(s)B/V_s<p_s,~\forall s'<s:\lambda(s)B/V_{s'}\geq p_{s'}\}.$$
\begin{prop}\label{prop:firststepFLCI}
$\mathcal{S}_{\text{ci}}\ne \emptyset$, and $\exists s_{\text{ci}}\in\mathcal{S}_{\text{ci}}:~w^{\text{CI}}_{M,s}=p_s1\{s<s_{\text{ci}}\}+\lambda(s_{\text{ci}})B/V_s1\{s\geq s_{\text{ci}}\}$.
\end{prop}
Then, $\bm{w}^{\text{CI}}_M$ can be found by evaluating $\sigma^2(\bm{w})$ at at most $S$ candidate values. The weights in Proposition \ref{prop:firststepFLCI} are similar to those in Theorem \ref{thm_minimax}: they do not downweight the precisely estimated CATEs and they downweight the less precise ones by assigning them a weight proportional to one over their variance. Next, $\bm{w}^{\text{CI}}$ can be found: 1) by minimizing $Q_{1-\alpha}\left(M,\sigma(\bm{w}^{\text{CI}}_M)\right)$ across $M \in (0,B)$, a univariate minimization problem, that can be solved via a grid search; 2) comparing the result to $Q_{1-\alpha}\left(0,\sigma(\bm{p})\right)$ and $Q_{1-\alpha}\left(B,0\right)$.

\medskip
The confidence interval
$\widehat{\tau}(\bm{w}^{\text{CI}})+/-Q_{1-\alpha}\left(\overline{|B|}(\bm{w}^{\text{CI}}),\sigma(\bm{w}^{\text{CI}})\right)$
relies on Assumption \ref{hyp:CATE_hat_normal}, a parametric normality assumption. I conjecture that, as shown by \cite{armstrong2018optimal} and \cite{armstrong2021sensitivity} for similar minimax CIs in different contexts, without that assumption this CI remains valid when the sample size goes to infinity, provided one assumes that Assumption \ref{hyp:CATE_hat_p4} holds with a tuning parameter going to zero when the sample size goes to infinity. While I do not derive such asymptotic results, in Section \ref{sec:simulations} I find that the CI I propose has close-to-nominal coverage in simulations ran in the data used in my first application. Thus, it can have good coverage in realistic settings where Assumption \ref{hyp:CATE_hat_normal} may fail.

\subsection{Assuming that CATEs are bounded and are all of the same, known sign}

Under Assumption \ref{hyp:CATE_hat_sign}, 
$\tau$ is assumed to be positive.
Then, it might make sense to consider a confidence lower bound (CB) instead of a CI for $\tau$.
Letting $\overline{B}(\bm{w})=B\sum_{s=1}^S(w_s-p_s)_+$ and $\underline{B}(\bm{w})=B\sum_{s=1}^S(w_s-p_s)_-$, under Assumption \ref{hyp:CATE_hat_p4}
\begin{equation}\label{eq:boundbias_1}
\underline{B}(\bm{w})\leq B(\bm{w}) \leq \overline{B}(\bm{w}).
\end{equation}
Then, letting $q_p$ denote the quantile of order $p$ of a standard normal distribution, under Assumption \ref{hyp:CATE_hat_normal} we have that for any $\alpha \in (0,1)$,
\begin{align*}
1-\alpha=&P\left(\widehat{\tau}(\bm{w})-\tau \leq B(\bm{w})+\sigma(\bm{w})z_{1-\alpha} \right)\\
\leq & P\left(\widehat{\tau}(\bm{w})-\tau \leq \overline{B}(\bm{w})+\sigma(\bm{w})z_{1-\alpha} \right).
\end{align*}
Therefore,
$$\left[\widehat{\tau}(\bm{w})-\overline{B}(\bm{w})-\sigma(\bm{w})z_{1-\alpha},+\infty\right]$$
is a one-sided CI with coverage larger than $1-\alpha$ for all $B(\bm{w})$.
The expected excess length (EEL) of its CB is
\begin{align*}
&E\left(\tau-(\widehat{\tau}(\bm{w})-\overline{B}(\bm{w})-\sigma(\bm{w})z_{1-\alpha})\right)\\
=&-B(\bm{w})+\overline{B}(\bm{w})+\sigma(\bm{w})z_{1-\alpha}\\
\leq& \overline{\text{EEL}}(\bm{w}):=\overline{B}(\bm{w})-\underline{B}(\bm{w})+\sigma(\bm{w})z_{1-\alpha}=B\sum_{s=1}^S|w_s-p_s|+\sigma(\bm{w})z_{1-\alpha}.
\end{align*}
Following \cite{armstrong2018} and \cite{armstrong2021sensitivity}, I propose to use the CB with minimax EEL, namely $\widehat{\tau}(\bm{w}^{\text{CB}})-\overline{B}(\bm{w}^{\text{CB}})-\sigma(\bm{w}^{\text{CB}})z_{1-\alpha}$, where $\bm{w}^{\text{CB}}$ is the minimizer of $\overline{\text{EEL}}(\bm{w})$ across all $\bm{w}\in \mathbb{R}^S:w_1\geq p_1$. Here, we restrict attention to weights such that $w_1\geq p_1$: the minimizer across the unrestricted set of weights does not always satisfy this condition, which comes with undesirable consequences as explained in the next section. Using arguments similar to those in Footnote \ref{footnote_boundsoptimalweights}, one can show that $w^{\text{CB}}_s\leq p_s$. Therefore, $w^{\text{CB}}_1= p_1$.
Letting $\bm{w}_{-1}=(w_2,...,w_S)$, the vector with coordinates $2$ to $S$ of $\bm{w}^{\text{CB}}$ is the minimizer of
$$B\sum_{s=1}^S(p_s-w_s)+z_{1-\alpha}\sqrt{\sum_{s=1}^Sw_s^2V_s}$$ across $\mathcal{R}^{\text{CB}}:=\{\bm{w}_{-1}:w_s\leq p_s\}.$ For any $k\in \{1,...,S\}$, let
\begin{equation*}
\sigma^2(k)=\frac{\sum_{s=1}^{k-1}p^2_sV_s}{1-\frac{B^2}{z^2_{1-\alpha}}\sum_{s=k}^{S}1/V_s},
\end{equation*}
and let
$$\mathcal{S}_{\text{cb}}=\{s:\sigma^2(s)>0,~\sigma^2(s) B/(z_{1-\alpha}V_{s})<p_{s},~\forall s'<s:\sigma^2(s) B/(z_{1-\alpha}V_{s'})\geq p_{s'}\}.$$
\begin{thm}\label{thm:CB}
If $p_SV_S\leq \sigma(\bm{p}) B/z_{1-\alpha}$, $\bm{w}^{\text{CB}}=\bm{p}$.
If $p_SV_S> \sigma(\bm{p}) B/z_{1-\alpha}$,
$\mathcal{S}_{\text{cb}}$ is not empty, and $\exists s_{\text{cb}}\in\mathcal{S}_{\text{cb}}:~w^{\text{CB}}_{s}=p_s1\{s<s_{\text{cb}}\}+\sigma(s_{\text{cb}}) B/(z_{1-\alpha}V_s)1\{s\geq s_{\text{cb}}\}\}$.
\end{thm}
Then, $\bm{w}^{\text{CB}}$ can be found by evaluating $\overline{\text{EEL}}(\bm{w})$ at (at most) $S$ candidate values. Interestingly, one may have $\bm{w}^{\text{CB}}=\bm{p}$.
When $\bm{w}^{\text{CB}}\ne \bm{p}$, the weights in Theorem \ref{thm:CB} are similar to those in Theorem \ref{thm_minimax}: they do not downweight precisely estimated CATEs and they downweight less precise ones with a weight proportional to one over their variance.

\section{Power-aware minimax estimators and confidence intervals}\label{sec:power}

For any $\bm{w}$, let
$$\text{CI}_{1-\alpha}(\bm{w})=\left[\widehat{\tau}(\bm{w})-Q_{1-\alpha}\left(\overline{|B|}(\bm{w}),\sigma(\bm{w})\right),\widehat{\tau}(\bm{w})+Q_{1-\alpha}\left(\overline{|B|}(\bm{w}),\sigma(\bm{w})\right)\right]$$
be the bias-aware $1-\alpha$-level CI for $\tau$ attached to $\widehat{\tau}(\bm{w})$ under Assumptions \ref{hyp:CATE_hat}, \ref{hyp:CATE_hat_p4}, and \ref{hyp:CATE_hat_normal}.

\subsection{Necessary condition for admissible tests}

\begin{prop}\label{prop:conditions_under_whichpowerlower}
Assume that Assumptions \ref{hyp:CATE_hat}, \ref{hyp:CATE_hat_p4}, and \ref{hyp:CATE_hat_normal} hold. For any $\bm{w}:\sum_{s=1}^S w_s> 0,~w_s<p_s~\forall s$, for any $\lambda \in \left(1,\underset{s\in \{1,...,S\}}{\min}p_s/w_s\right]$, $$P(0\in \text{CI}_{1-\alpha}(\bm{w}))>P(0\in \text{CI}_{1-\alpha}(\lambda\bm{w})).$$
\end{prop}
Proposition \ref{prop:conditions_under_whichpowerlower} shows that for any linear estimator that strictly downweights all CATEs ($w_s<p_s~\forall s$), one can construct an alternative estimator that will lead to tests of the null that $\tau=0$ that control size, and that have strictly higher power whenever $\tau \ne 0$. Thus, linear estimators that strictly downweight all CATEs lead to inadmissible tests. While under Assumption \ref{hyp:CATE_hat_p4}, the minimax-linear estimator of $\tau$ does not strictly downweight all CATEs, the minimax-linear estimator of $\tau$ under Assumption \ref{hyp:CATE_hat_het} can strictly downweight all CATEs. This arises when the conditions of Point 3 of Theorem \ref{thm_minimax2} are met, in which case the minimax-linear estimator downnweights all CATEs by the same factor. Then, letting $\lambda=p_s/w^H_s$, Proposition \ref{prop:conditions_under_whichpowerlower} implies that tests of $\tau=0$ based on the unbiased estimator are uniformly more powerful than tests based on the minimax-linear estimator. Even when the conditions of Point 3 of Theorem \ref{thm_minimax2} are not met, one may still have that the minimax-linear estimator downnweights all CATEs, by different factors. This for instance arises in my second empirical application.

\medskip
This motivates considering the linear combination of CATE estimators that minimizes worst-case MSE, CI length, or CB EEL, across all linear combinations that do not strictly downweight all estimators. $\widehat{\tau}(\bm{w}^{\text{M}}_{\text{B}})$, $\widehat{\tau}(\bm{w}^{\text{CI}})$, and $\widehat{\tau}(\bm{w}^{\text{CB}})$ automatically satisfy this criterion. $\widehat{\tau}(\bm{w}^H)$ may not, so under Assumption \ref{hyp:CATE_hat_het}, instead of $\widehat{\tau}(\bm{w}^H)$ one may consider $\widehat{\tau}(\bm{w}^P)$, where
$$\bm{w}^{P}=\argmin_{\bm{w}\in \mathbb{R}^S_+:w_{(1)}\geq 1/S} \overline{\text{MSE}}^H(\bm{w}).$$
Lemma \ref{lem_worst-case-MSE_boundhet}, Point 2 of Lemma \ref{lem_average_opt_weights_lowerthanone}, and Lemma \ref{lem_getting_rid_max} (used in the proof of Lemma \ref{lem_average_opt_weights_lowerthanone}) apply to any $\bm{w}\in \mathbb{R}^S_+$, so they of course apply to any $\bm{w}\in \mathbb{R}^S_+:w_{(1)}\geq 1/S$. One can check that Point 1 of Lemma \ref{lem_average_opt_weights_lowerthanone}\footnote{The beginning of the proof of Case 1 needs to be modified as follows: ``Assume that $s\mapsto w^P_{[s]}$ is not constant, and $\sum_{s=1}^{S}w^P_{[s]}>1.$ We cannot have $w^P_{[s]}=0$ for all $s$, and let $s_0:\max\{s:w^P_{[s]}>0\}$. If $\sum_{s=1}^{S}w^P_{[s]}>1,$, we must have $w^P_{[1]}>1/S$.
Then, for a strictly positive $$\eps<\min\left(\sum_{s=1}^{S}w^P_{[s]}-1,w^P_{[s_0]}S,w^P_{[1]}-1/S\right),$$ let $\tilde{w}_{[s]}=w^P_{[s]}-\eps/S$ if $s\leq s_0$, $\tilde{w}_{[s]}=w^P_{[s]}=0$ otherwise. $\tilde{w}_{[1]}>1/S$, $\tilde{w}_{[s_0]}=w^P_{[s_0]}-\eps/S>0$. Therefore, $\bm{\tilde{w}}\in \{\bm{w}\in \mathbb{R}^S_+:w_{(1)}\geq 1/S\}$, and $s\mapsto \tilde{w}_{[s]}$ is decreasing.'' The rest of the proof follows.} and Lemma \ref{lem_permutation} still apply to $\bm{w}^P$. Moreover, it is easy to show by contradiction that $w_{[1]}^P=1/S$. Then, a similar result as that in Point 1 of Theorem \ref{thm_minimax2} follows:
$\bm{w}^P=\argmin_{\bm{w}\in \mathcal{R}^P} \overline{\text{MSE}}^{H,d}(\bm{w})$, where
$$\mathcal{R}^P:=\left\{\bm{w}\in \mathbb{R}^S:w_1=1/S,~\forall
s \in \{1,...,S-1\}:w_{s+1}-w_{s}\leq 0,~-w_S\leq 0\right\}.$$
Then, $\bm{w}^H$ can be approximated by standard convex quadratic programming algorithms.

\subsection{Further necessary condition for admissible tests, when minimax weights and CATEs are uncorrelated}

\begin{prop}\label{prop:conditions_under_whichpowerlower2}
Assume that Assumptions \ref{hyp:CATE_hat}, \ref{hyp:CATE_hat_p4}, and \ref{hyp:CATE_hat_normal} hold.
For any $\bm{w}:\sum_{s=1}^S w_s>0$, if $\sum_{s=1}^S p_s \frac{w_s}{p_s}\tau_s=\sum_{s=1}^S p_s \frac{w_s}{p_s}\times \sum_{s=1}^S p_s \tau_s$,
$$\sum_{s=1}^S w_s < \sigma(\bm{w})/\sigma(\bm{p})\Rightarrow P(0\in \text{CI}_{1-\alpha}(\bm{w}))>P(0\in \text{CI}_{1-\alpha}(\bm{p})).$$
\end{prop}
Proposition \ref{prop:conditions_under_whichpowerlower2} implies that if a minimax estimator, CI, or CB is such that $\sum_{s=1}^S w_s < \sigma(\bm{w})/\sigma(\bm{p})$, it can again lead to inadmissible tests, if
$\sum_{s=1}^S p_s \frac{w_s}{p_s}\tau_s=\sum_{s=1}^S p_s \frac{w_s}{p_s}\times \sum_{s=1}^S p_s \tau_s$. This condition requires that the CATEs, $\tau_s$, and the extent to which they are downweighted by the minimax procedure, $w_s/p_s$, are uncorrelated. This for instance holds if the CATEs are independent of $(p_s,V_s)_{s\in \{1,...,S\}}$. It also mechanically holds if the CATEs are constant.

\medskip
While here, inadmissibility only holds under a strong no-correlation condition, if one is not ready to assume ex-ante that this condition fails, one may want to consider even more ``power-aware'' minimax-linear procedures, namely the linear combination of CATE estimators that minimizes worst-case MSE, CI length, or CB EEL, across all linear combinations such that $\exists s: w_s\geq p_s,\sum_{s=1}^S w_s \geq  \sigma(\bm{w})/\sigma(\bm{p})$. In my empirical applications, $\widehat{\tau}(\bm{w}^{\text{M}}_{\text{B}})$ and $\widehat{\tau}(\bm{w}^{\text{CI}})$ are such that $\sum_{s=1}^S w_s \geq  \sigma(\bm{w})/\sigma(\bm{p})$, but there may be other applications where those estimators do not automatically satisfy that condition.

\medskip
Instead of requiring $\exists s: w_s\geq p_s,\sum_{s=1}^S w_s \geq  \sigma(\bm{w})/\sigma(\bm{p})$, one could restrict the minimization set to linear combinations such that $\sum_{s=1}^S w_s=1$, a stronger condition.\footnote{The solution of both minimization problems must be such that $\sigma(\bm{w})/\sigma(\bm{p})\leq 1$, as otherwise $\bm{p}$ would dominate.} However, while this stronger criterion would have the benefit of ensuring that minimax estimators are unbiased if the treatment effect is constant, it is less-well grounded in statistical decision theory: for now, I have not been able to show that, under reasonable assumptions, a minimax estimator is inadmissible when $\sigma(\bm{w})/\sigma(\bm{p})\leq \sum_{s=1}^S w_s<1$.

\subsection{Recommendations}

While requiring $\exists s: w_s\geq p_s,\sum_{s=1}^S w_s \geq  \sigma(\bm{w})/\sigma(\bm{p})$ ensures that the minimax procedures I study do not lead to inadmissible tests, these conditions do not ensure that those tests are maximin, or just that they have more power than tests based on the unbiased estimator. \eqref{eq:power} in the proofs shows that the probability that zero belongs to the CI attached to $\widehat{\tau}(\bm{w})$ is equal to
$$\Phi\left(-\frac{E\left(\widehat{\tau}(\bm{w})\right)}{\sigma(\bm{w})}+Q_{1-\alpha}\left(\overline{|B|}(\bm{w})/\sigma(\bm{w}),1\right)\right)-\Phi\left(-\frac{E\left(\widehat{\tau}(\bm{w})\right)}{\sigma(\bm{w})}-Q_{1-\alpha}\left(\overline{|B|}(\bm{w})/\sigma(\bm{w}),1\right)\right).$$
Then, I recommend that researchers using minimax-length CIs compute
$$\frac{1-\left(\Phi\left(-\frac{\widehat{\tau}(\bm{w})}{\sigma(\bm{w})}+Q_{1-\alpha}\left(\overline{|B|}(\bm{w})/\sigma(\bm{w}),1\right)\right)-\Phi\left(-\frac{\widehat{\tau}(\bm{w})}{\sigma(\bm{w})}-Q_{1-\alpha}\left(\overline{|B|}(\bm{w})/\sigma(\bm{w}),1\right)\right)\right)
}{1-\left(\Phi\left(-\frac{\widehat{\tau}(\bm{p})}{\sigma(\bm{p})}+Q_{1-\alpha}\left(0,1\right)\right)-\Phi\left(-\frac{\widehat{\tau}(\bm{p})}{\sigma(\bm{p})}-Q_{1-\alpha}\left(0,1\right)\right)\right)},$$
an estimator of the power gain induced by a minimax estimator or CI. Moreover, I recommend that researchers conduct simulations in their own data to assess if minimax procedures can lead to power gains in a controlled setting closely mimicking that they consider. In Section \ref{sec:simulations}, I run such simulations.

\section{Applications}\label{sec:application}

\subsection{\cite{behaghel2017ready}}

The authors conducted an SRCT to estimate the effect of a boarding school for disadvantaged students in France. The boarding school's pedagogy is similar to that of ``No Excuse'' charter schools in the US. It has capacity constraints at the gender $\times$ grade level, and in 2009 and 2010, the school had more applicants than seats in 14 gender $\times$ grade strata. In each stratum, seats were randomly offered to some applicants.\footnote{Here, I do not take into account the fact that randomization followed a waitlist process, which generates complications orthogonal to the issues discussed in this paper. This explains why some numbers below do not exactly match the corresponding numbers in \cite{behaghel2017ready}.} The probability of receiving a treatment offer varies substantially across strata: it ranges from 0.17 to 0.93. Two years after the randomization, 363 applicants out of the 395 that participated in a lottery took a standardized maths test. Those students are the study sample. The main outcome in \cite{behaghel2017ready}, and the sole outcome in this re-analysis, is students' maths test score, two years after the lottery, divided by the standard deviation of the tests scores of control students in their strata. $\tau_s$ is the intention to treat effect of receiving an offer to join the school in stratum $s$. Two years after the lottery, the first-stage effect of receiving an offer on the number of years spent in the school is equal to 1.34, so the $\tau_s$s are effects of having spent 1.34 years in the boarding school.\footnote{The first-stage effects may vary across strata, which would complicate this interpretation. However, I use the \st{multi\_site} Stata package of \cite{de2024estimating} to estimate the variance of first-stage effects across strata, and I cannot reject the null that all first-stages are equal (t-stat=1.10).}

\medskip
An abundant literature has estimated ATEs and CATEs of similar interventions, expressed in percentage points of the control group's standard deviation $\sigma_0$. Based on the literature, $0.5\sigma_0$ is a plausible upper bound for the effect of spending one year in the boarding school studied by \cite{behaghel2017ready}. The paper studying the closest intervention is \cite{curto2014potential}, who study a ``No Excuse'' charter boarding school in Washington DC. In their full sample, they find that one year spent in the school increases students' math test scores by 0.23$\sigma_0$. They also estimate CATEs in eight subgroups of students: males/females, students benefiting/not benefiting from the free lunch program, students in/not in special education, and students above/below the median at baseline. The estimated effects in those 8 subgroups are included between $0.04$ and $0.36\sigma_0$, and in 7 of the 8 subgroups one can reject at the 90\% level that the effect is greater than $0.5\sigma_0$, the only exception being the special education stratum that only has 30 students. Results from \cite{angrist2010inputs}, \cite{dobbie2011high}, and \cite{abdulkadirouglu2011accountability}, three papers studying successful non-boarding ``No Excuse'' charter schools in New-York and Boston, also suggest that $0.5\sigma_0$ is a plausible upper bound. Together, these papers estimate 14 CATEs of spending one year in those schools on students' math test scores. All estimates are included between $0.18$ and $0.36\sigma_0$, and for 13 of the 14 CATEs, one can reject at the 90\% level an effect greater than $0.5\sigma_0$.
Accordingly, I assume that Assumption \ref{hyp:CATE_hat_p4} holds, with $B=1.34\times 0.5$. As a robustness check, I will also let $B=1.34\times 0.6$.

\medskip
Results are shown in Table \ref{table_results}. Estimators are computed as described in the previous sections, and for any $\bm{w}$ I let
$\widehat{V}\left(\widehat{\tau}(\bm{w})\right)=\sum_{s=1}^Sw_s^2 \widehat{V}_s,$ where $\widehat{V}_s:=1/n_{0,s}+\widehat{\sigma}_{1,s}/n_{1,s}$ is the usual robust estimator of the variance of the ITT estimator in stratum $s$, given that by construction $\widehat{\sigma}_{0,s}=1$. $\widehat{\tau}(\bm{w}^{M}_{0.5})$ downweights the three strata with the least precisely estimated ITTs, and $\sum_{s=1}^{14}w^{M}_{0.5,s}=0.944$. The first two columns of Table \ref{table_results} show that such shrinkage does not seem to lead to a bias, but reduces variance:  $\widehat{\tau}(\bm{p})=$ 0.278 (s.e.=0.134), $\widehat{\tau}(\bm{w}^{M}_{0.5})=$0.268 (s.e.=0.122), so the two estimators are very close but the standard error of $\widehat{\tau}(\bm{w}^{M}_{0.5})$ is 9.2\% smaller.

\medskip
Yet, this precision gain may be offset by $\widehat{\tau}(\bm{w}^{M}_{0.5})$'s bias. As
$$\left(E\left(\widehat{\tau}(\bm{p})\right)-E\left(\widehat{\tau}(\bm{w}^{M}_{0.5})\right)\right)^2=E\left((\widehat{\tau}(\bm{p})- \widehat{\tau}(\bm{w}^{M}_{0.5}))^2\right)-V\left(\widehat{\tau}(\bm{p})- \widehat{\tau}(\bm{w}^{M}_{0.5})\right),$$
to estimate the square bias of $\widehat{\tau}(\bm{w}^{M}_{0.5})$ I use
$$\max\left((\widehat{\tau}(\bm{p})- \widehat{\tau}(\bm{w}^{M}_{0.5}))^2-\widehat{V}\left(\widehat{\tau}(\bm{p})- \widehat{\tau}(\bm{w}^{M}_{0.5})\right),0\right),$$
where $$\widehat{V}\left(\widehat{\tau}(\bm{p})- \widehat{\tau}(\bm{w}^{M}_{0.5})\right)=\sum_{s=1}^S\left(w^{M}_{0.5,s}-p_s\right)^2 \widehat{V}_s.$$
As $\widehat{\tau}(\bm{p})$ and $\widehat{\tau}(\bm{w}^{M}_{0.5})$ are extremely close, $(\widehat{\tau}(\bm{p})- \widehat{\tau}(\bm{w}^{M}_{0.5}))^2-\widehat{V}\left(\widehat{\tau}(\bm{p})- \widehat{\tau}(\bm{w}^{M}_{0.5})\right)<0$, so the fourth line of the table shows that the estimated RMSE of $\widehat{\tau}(\bm{w}^{M}_{0.5})$ is again 9.2\% smaller than that of $\widehat{\tau}(\bm{p})$. Rather than comparing the estimated RMSE of the two estimators, one can compare their worst-case RMSE under Assumption \ref{hyp:CATE_hat_p4}. The fifth line of the table shows that the worst-case RMSE of $\widehat{\tau}(\bm{w}^{M}_{0.5})$ is 5\% smaller than that of $\widehat{\tau}(\bm{p})$.

\medskip
The third column of Table \ref{table_results} shows that results are fairly robust to letting $B=1.34\times 0.6$: the estimated RMSE and worst-case RMSE of $\widehat{\tau}(\bm{w}^{M}_{0.6})$ are respectively 7.1 and 4.1\% smaller than that of $\widehat{\tau}(\bm{p})$. The fourth column shows that assuming that $B=1.34\times 0.5$ and that the outcome is homoscedastic also does not greatly change the results, though the gain in terms of worst-case RMSE becomes smaller:
the estimated RMSE and worst-case RMSE of $\widehat{\tau}(\bm{w}^{\text{hom}}_{0.5})$ are respectively 7.5 and 1.9\% smaller than that of $\widehat{\tau}(\bm{p})$. While the other variance estimators shown in the table do not account for the fact that the weights are estimated, the variance estimator of $\widehat{\tau}(\bm{w}^{\text{hom}}_{0.5})$ does not suffer from this issue, as the weights attached to that estimator do not need to be estimated.

\medskip
Turning to inference, the fifth column of Table \ref{table_results}  shows $\widehat{\tau}(\bm{w}^{\text{CI}})$ and $$\left[\widehat{\tau}(\bm{w}^{\text{CI}})-Q_{0.95}\left(\overline{|B|}(\bm{w}^{\text{CI}}),\sigma(\bm{w}^{\text{CI}})\right),\widehat{\tau}(\bm{w}^{\text{CI}})+Q_{0.95}\left(\overline{|B|}(\bm{w}^{\text{CI}}),\sigma(\bm{w}^{\text{CI}})\right)\right],$$ computed for $B=1.34\times 0.5$, following the steps outlined in the previous section. The minimax fixed-length 95\%-level CI is equal to [0.021,0.519], and its length is 5.2\% smaller than that of the CI attached to $\widehat{\tau}(\bm{p})$ ([0.015,0.540]).

\medskip
Finally, turning to power, $\sum_{s=1}^S w_s$ is always larger than the ratio of the standard errors of the minimax and unbiased estimators, so the minimax estimator is not inadmissible, and I estimate that $\widehat{\tau}(\bm{w}^{\text{CI}})$ leads to a 4.3\% power gain relative to $\widehat{\tau}(\bm{p})$.

\medskip
Overall, minimax estimators and CI seem to lead to a small but not completely negligible precision gain with respect to $\widehat{\tau}(\bm{p})$, included between 5 and 10\% for most pairs of metrics and estimators I consider. To put this into perspective, to achieve a 5\% precision gain by controlling for covariates (or additional covariates if some covariates are already controlled for), the covariates need to explain $=1-0.95^2\approx$10\% of the outcome's variance \cite[see, e.g., Section 5.2 of][]{athey2017econometrics}.\footnote{Of course, minimax estimators and controlling for covariates are not mutually exclusive strategies to improve precision, they can be combined.} Another way to benchmark this precision gain is to compare it to that obtained from a regression of test scores on a treatment offer and strata fixed effects. This yields a variance-weighted average of CATE estimators \cite[][]{Angrist08}, which often has a lower variance than $\widehat{\tau}(\bm{p})$ and is in fact the best linear unbiased estimator of $\tau$ if the CATEs are homogenous and the outcome is homoscedastic ($\sigma_{1,s}=1$ for all $s$). The corresponding estimator is equal to 0.257. Its standard error (0.133), is only 0.9\% smaller than that of $\widehat{\tau}(\bm{p})$: in this application minimax estimators seem to be more precise than the strata fixed-effects estimator.


\begin{table}[H]
\begin{center}
\caption{Minimax estimators and CI in \cite{behaghel2017ready}}
\begin{tabular}{l c c c c c}
\hline \hline
 & $\widehat{\tau}(\bm{p})$ & $\widehat{\tau}(\bm{w}^{M}_{0.5})$ & $\widehat{\tau}(\bm{w}^{M}_{0.6})$ & $\widehat{\tau}(\bm{w}^{\text{hom}}_{0.5})$ & $\widehat{\tau}(\bm{w}^{\text{CI}})$   \\
\hline
Point estimate & 0.278 & 0.268  & 0.270 & 0.273 & 0.270  \\
Robust s.e. & 0.134 & 0.122  & 0.124 & 0.124 & \\
Robust s.e./Robust s.e.$\left(\widehat{\tau}(\bm{p})\right)$ & 1 & 0.908 & 0.929 &  0.925 & \\
$\widehat{\text{RMSE}}/\widehat{\text{RMSE}}\left(\widehat{\tau}(\bm{p})\right)$  & 1 & 0.908 & 0.929 & 0.925 & \\
$\widehat{\overline{\text{RMSE}}}/\widehat{\overline{\text{RMSE}}}\left(\widehat{\tau}(\bm{p})\right)$  & 1 & 0.950 & 0.959 & 0.981 & \\
95\% level CI  & [0.015,0.540] &  & &  & [0.021,0.519] \\
$\sum_{s=1}^S w_s$ & 1 & 0.944  & 0.960 & 0.934  & 0.950 \\
$\widehat{\text{POWER}}/\widehat{\text{POWER}}\left(\widehat{\tau}(\bm{p})\right)$  & 1 & & & & 1.043 \\
\hline \hline
\end{tabular}\label{table_results}
\end{center}
\footnotesize{Notes: This table shows $\widehat{\tau}(\bm{p})$, $\widehat{\tau}(\bm{w}^{M}_{0.5})$, $\widehat{\tau}(\bm{w}^{M}_{0.6})$, $\widehat{\tau}(\bm{w}^{\text{hom}}_{0.5})$, and $\widehat{\tau}(\bm{w}^{\text{CI}})$ in \cite{behaghel2017ready}. The treatment is defined as being offered a seat in the boarding school. The outcome is students' standardized maths test scores two years after the lottery.}
\end{table}

\subsection{\cite{connors1996effectiveness}}

In this section, I revisit \cite{connors1996effectiveness}, a matching study also revisited by \cite{crump2009}. The authors study the impact of right heart catheterization (RHC) on patient mortality. RHC is a diagnostic procedure used for critically-ill patients. The data contain information on 5,735 patients. For each patient, I observe the treatment status $D_s$, defined as RHC being applied within 24 hours of admission, the outcome $Y_s$, an indicator for survival at 30
days, and 71 covariates deemed related to the decision to
perform the RHC by a panel of experts. Using a propensity score matching approach, the authors concluded that RHC causes a substantial increase in patient mortality.

\medskip
Column (1) of Table \ref{table_results_connors} shows $\widehat{\tau}(\bm{p})$, the feasible AIPW estimator where $e(X_s)$, $\mu_0(X_s)$, and $\mu_1(X_s)$ are replaced by their estimators. I follow \cite{hirano2001estimation} and \cite{crump2009}, and estimate the propensity score $e(X_s)$ using a logistic regression that includes all the covariates.
As the outcome is binary, I also use logistic regressions including all covariates to estimate $\mu_0(X_s)$ and $\mu_1(X_s)$. Then, I let $$\widehat{V}_s=\widehat{\mu}_0(X_s)(1-\widehat{\mu}_0(X_s))/(1-\widehat{e}(X_s))+\widehat{\mu}_1(X_s)(1-\widehat{\mu}_1(X_s))/\widehat{e}(X_s).$$
Column (2) of Table \ref{table_results_connors} shows the minimax-linear estimator $\widehat{\tau}(\bm{w}^M_{0.2})$. $0.2$ is a large, rarely seen effect size for a binary outcome. $\widehat{\tau}(\bm{w}^M_{0.2})$ downweights 297 patients, namely 5.2\% of the sample, and $\sum_{s=1}^{5735}w^{M}_{0.2,s}=0.977$. Such shrinkage does not seem to lead to a bias, but reduces variance:  $\widehat{\tau}(\bm{p})=$-0.064 (s.e.=0.016) and $\widehat{\tau}(\bm{w}^{M}_{0.5})=$-0.065 (s.e.=0.014), so the two estimators are very close but the standard error of $\widehat{\tau}(\bm{w}^{M}_{0.2})$ is 10.7\% smaller.
This precision gain is not entirely offset by $\widehat{\tau}(\bm{w}^{M}_{0.2})$'s bias.
As $\widehat{\tau}(\bm{p})$ and $\widehat{\tau}(\bm{w}^{M}_{0.2})$ are extremely close, $(\widehat{\tau}(\bm{p})- \widehat{\tau}(\bm{w}^{M}_{0.2}))^2-\widehat{V}\left(\widehat{\tau}(\bm{p})- \widehat{\tau}(\bm{w}^{M}_{0.2})\right)<0$, so the fourth line of the table shows that the estimated RMSE of $\widehat{\tau}(\bm{w}^{M}_{0.2})$ is again 10.7\% smaller than that of $\widehat{\tau}(\bm{p})$. The fifth line of the table shows that the worst-case RMSE of $\widehat{\tau}(\bm{w}^{M}_{0.2})$ is 6\% smaller than that of $\widehat{\tau}(\bm{p})$.

\medskip
The third column of Table \ref{table_results_connors} shows that results are fairly robust to letting $B=0.3$: the estimated RMSE and worst-case RMSE of $\widehat{\tau}(\bm{w}^{M}_{0.3})$ are respectively 7.9 and 4.2\% smaller than that of $\widehat{\tau}(\bm{p})$.
Turning to inference, the fourth column of Table \ref{table_results_connors}  shows $\widehat{\tau}(\bm{w}^{\text{CI}})$ and its 95\%-level confidence interval for $B=0.2$. The minimax fixed-length 95\%-level CI is equal to [-0.093, -0.036], and its length is 6.6\% smaller than that of the CI attached to $\widehat{\tau}(\bm{p})$ ([-0.095,-0.033]).
Finally, turning to power, $\sum_{s=1}^S w_s$ is always larger than the ratio of the standard errors of the minimax and unbiased estimators, so the minimax estimator is not inadmissible. I also estimate that relative to $\widehat{\tau}(\bm{p})$, $\widehat{\tau}(\bm{w}^{\text{CI}})$ leads to a 1.2\% increase in power to reject the null that $\tau=0$. The power gain is small, because estimated power is already very large with the unbiased estimator (>98\%).

\medskip
Again, minimax estimators and CI seem to lead to a small but not completely negligible precision gain with respect to $\widehat{\tau}(\bm{p})$. As a benchmark, the trimming estimator of \cite{crump2009} is equal to -0.069 and its standard error (0.014) is only 0.5\% smaller than that of $\widehat{\tau}(\bm{w}^{M}_{0.2})$. Thus, $\widehat{\tau}(\bm{w}^{M}_{0.2})$ is nearly as precise as the trimming estimator, without changing the goalpost.
\begin{table}[H]
\begin{center}
\caption{Minimax estimators and CI in \cite{connors1996effectiveness}}
\begin{tabular}{l c c c c}
\hline \hline
 & $\widehat{\tau}(\bm{p})$ & $\widehat{\tau}(\bm{w}^{M}_{0.2})$ & $\widehat{\tau}(\bm{w}^{M}_{0.3})$ & $\widehat{\tau}(\bm{w}^{\text{CI}})$   \\
\hline
Point estimate & -0.064  & -0.065   & -0.066  & -0.064   \\
Robust s.e. & 0.016  & 0.014  & 0.015  & \\
Robust s.e./Robust s.e.$\left(\widehat{\tau}(\bm{p})\right)$ & 1 & 0.893  & 0.921 & \\
$\widehat{\text{RMSE}}/\widehat{\text{RMSE}}\left(\widehat{\tau}(\bm{p})\right)$   & 1 & 0.893  & 0.921 & \\
$\widehat{\overline{\text{RMSE}}}/\widehat{\overline{\text{RMSE}}}\left(\widehat{\tau}(\bm{p})\right)$  & 1 & 0.940 & 0.958 & \\
95\% level CI  & [-0.095,-0.033] &  &  & [-0.093, -0.036] \\
$\sum_{s=1}^S w_s$ & 1 & 0.977   & 0.986  & 0.980   \\
$\widehat{\text{POWER}}/\widehat{\text{POWER}}\left(\widehat{\tau}(\bm{p})\right)$  & 1 & & & 1.012 \\
\hline \hline
\end{tabular}\label{table_results_connors}
\end{center}
\footnotesize{Notes: This table shows $\widehat{\tau}(\bm{p})$, $\widehat{\tau}(\bm{w}^{M}_{0.2})$, $\widehat{\tau}(\bm{w}^{M}_{0.3})$, and $\widehat{\tau}(\bm{w}^{\text{CI}})$ in \cite{connors1996effectiveness}. The treatment is right heart catheterization and the outcome is survival at 30 days.}
\end{table}

Finally, I compare the efficiency gains obtained assuming that the CATEs are bounded, to the gains obtained assuming that their heterogeneity is bounded. If $\tau<0$, Assumption \ref{hyp:CATE_hat_het} requires that
$\tau_s\in [(B+1)\tau,-(B-1)\tau]$, an interval of length $2B|\tau|$. As $\widehat{\tau}(\bm{p})<0$ Assumption \ref{hyp:CATE_hat_p4} with tuning parameter equal to $B|\widehat{\tau}(\bm{p})|$ requires that $\tau_s\in [B\widehat{\tau}(\bm{p}),-B\widehat{\tau}(\bm{p})]$, an interval of length $2B|\widehat{\tau}(\bm{p})|$. Then, the two assumptions are comparable. For $B\in \{1,...,5\}$, the ratio of the standard errors of $\widehat{\tau}(\bm{w}^{P})$ and $\widehat{\tau}(\bm{p})$ are respectively equal to $0.865$, $0.900$, $0.916$, $0.923$, and $0.927$. By comparison, for $B\in \{1,...,5\}$, the ratio of the standard errors of $\widehat{\tau}(\bm{w}^{M}_{B|\widehat{\tau}(\bm{p})|})$ and $\widehat{\tau}(\bm{p})$ are respectively equal to $0.768$, $0.852$, $0.889$, $0.910$, and $0.925$. Thus, with comparable tuning parameters, Assumption \ref{hyp:CATE_hat_p4} seems to lead to slightly larger efficiency gains than Assumption \ref{hyp:CATE_hat_het}.\footnote{I do not report results for $\widehat{\tau}(\bm{w}^{H})$: for $B\in \{3,4,5\}$, all CATE estimators, even the most precise ones, are shrunk towards zero with that estimator.}

\section{Simulations}\label{sec:simulations}

I run simulations based on the data of \cite{behaghel2017ready}. First, I generate potential outcomes assuming that the treatment has no effect: $Y_{is}(0)=Y_{is}(1)=Y_{is}$. Then, I reallocate 1,000 times the treatment, following the same stratified randomization as in the paper. For each simulated randomization, I compute $\widehat{\tau}(\bm{p})$ and $\widehat{\tau}(\bm{w}^{\text{CI}})$, as well as their CIs. Results are shown in Panel A of Table \ref{table_sims}. 95\% CIs based on $\widehat{\tau}(\bm{p})$ and $\widehat{\tau}(\bm{w}^{\text{CI}})$ both have nominal coverage close to 95\%. Thus, while I do not prove their asymptotic validity, CIs attached to $\widehat{\tau}(\bm{w}^{\text{CI}})$ can still have good coverage in realistic settings where Assumption \ref{hyp:CATE_hat_normal} may fail. On average across the simulations, the length of the CI attached to $\widehat{\tau}(\bm{w}^{\text{CI}})$ is 17.7\% smaller than the length of the CI attached to $\widehat{\tau}(\bm{p})$. In Panel B, I show results from the same simulations, except that I let $Y_{is}(1)=Y_{is}(0)+\widehat{\tau}(\bm{p})$. Then, the CI attached to $\widehat{\tau}(\bm{w}^{\text{CI}})$ contains zero much less often than that attached to $\widehat{\tau}(\bm{p})$: the minimax-length CI leads to a substantial power gain.

\begin{table}[H]
\begin{center}
\caption{Simulations based on the data of \cite{behaghel2017ready}}
\begin{tabular}{l c c}
\hline \hline
 & $\widehat{\tau}(\bm{p})$ & $\widehat{\tau}(\bm{w}^{\text{CI}})$   \\
\hline
\textbf{Panel A}: $Y_{is}(1)-Y_{is}(0)=0$  & & \\
CI coverage & 0.926   & 0.945 \\
Power to reject $\tau=0$ & 0.074 & 0.055 \\
CI lenth / CI length $\left(\widehat{\tau}(\bm{p})\right)$ & 1 & 0.823 \\
 & & \\
\textbf{Panel B}: $Y_{is}(1)-Y_{is}(0)=\widehat{\tau}(\bm{p})$  & & \\
CI coverage & 0.926   & 0.945 \\
Power to reject $\tau=0$ & 0.456 & 0.625 \\
CI lenth / CI length $\left(\widehat{\tau}(\bm{p})\right)$ & 1 & 0.823 \\
\hline \hline
\end{tabular}\label{table_sims}
\end{center}
\end{table}

\section{Conclusion}

I derive minimax-linear estimators of, and CIs for, an average treatment effect (ATE) that can be decomposed as a weighted average of conditional average treatment effects (CATEs), under various restrictions on the CATEs. First, I assume that the magnitude of the CATEs is bounded. Then I assume that their heterogeneity is bounded. In two empirical applications, minimax-linear estimators and CIs lead to small but non-negligible precision gains. Minimax-linear estimators can sometimes lead to inadmissible tests of the null of no treatment effect. For instance, they can have uniformly less power than tests based on the unbiased estimator. I provide diagnostic checks researchers can use to assess if this is or not a concern in their application. Those diagnostics suggest that minimax estimators do not only lead to precision gains in the applications I revisit: they also lead to small but non-negligible power gains.

\newpage
\bibliography{biblio}

\newpage
\section*{Proofs}

\subsection*{Proof of Lemma \ref{lem_worst-case-MSE}}

\begin{align*}
E\left(\left(\widehat{\tau}(\bm{w})-\tau\right)^2\right)=&V(\widehat{\tau}(\bm{w}))+\left(E\left(\widehat{\tau}(\bm{w})\right)-\tau\right)^2\\
=&\sum_{s=1}^Sw^2_sV(\widehat{\tau}_s)+\left(\sum_{s=1}^S(w_s-p_s)\tau_s\right)^2\\
= &\sum_{s=1}^Sw_s^2V_s+\left(\sum_{s=1}^S(w_s-p_s)\tau_s\right)^2\\
\leq&\sum_{s=1}^Sw_s^2V_s+\left(\sum_{s=1}^S|w_s-p_s||\tau_s|\right)^2\\
\leq&\left(\sum_{s=1}^Sw^2_sV_s+B^2\left(\sum_{s=1}^S|w_s-p_s|\right)^2\right).
\end{align*}
The first equality follows from the fact that an estimator's MSE is the sum of its variance and squared bias. The second equality follows from the fact $\bm{w}$ is deterministic, from Equations \eqref{eq:lin_comb_CATE} and \eqref{eq:decomposition}, and from Point \ref{hyp:CATE_hat_p1} of Assumption \ref{hyp:CATE_hat}. The third equality follows from Point \ref{hyp:CATE_hat_p3} of Assumption \ref{hyp:CATE_hat}. The first inequality follows from the fact that for any real number $a$, $a^2=|a|^2$, from the triangle inequality, and from the fact that $x\mapsto x^2$ is increasing on $\mathbb{R}_+$. The second inequality follows from Assumption \ref{hyp:CATE_hat_p4}.
The sharpness of the upper bound follows from plugging ${\tau}_s= B\left(1\{w_s\geq p_s\}-1\{w_s<p_s\}\right)$ into the second equality in the previous display.

\subsection*{Proof of Theorem \ref{thm_minimax}}

In view of Lemma \ref{lem_worst-case-MSE}, to prove the result we need to prove that $$\underset{\bm{w}\in \mathbb{R}^S} {\argmin}\overline{\text{MSE}}(\bm{w})=\bm{w}^M_B.$$ With a slight abuse of notation, in the proof $\bm{w}^M_B$ refers to $\underset{\bm{w}\in \mathbb{R}^S} {\argmin}\overline{\text{MSE}}(\bm{w})$, and the proof amounts to showing that the solution of this minimization problem coincides with the expression for $\bm{w}^M_B$ given in the text.

First, assume that $\bm{w}^M_B$  has at least one coordinate that is strictly larger than the corresponding coordinate of $(p_1,...,p_S)$. Without loss of generality, assume that $w^M_{B,1}>p_1$. One has
$\overline{\text{MSE}}(\bm{w}^M_B)>\overline{\text{MSE}}(p_1,w^M_{B,2},...,w^M_{B,S})$, a contradiction. Therefore, each coordinate of $\bm{w}^M_B$ is at most as large as the corresponding coordinate of $(p_1,...,p_S)$. Accordingly, finding the minimax-linear estimator is equivalent to minimizing $\overline{\text{MSE}}(\bm{w})$  with respect to $\bm{w}$, across all $\bm{w}=(w_1,...,w_S)$ such that $w_s\leq p_s$ for all $s\in \{1,...,S\}$.

\medskip
If $w_s\leq p_s$ for all $s\in \{1,...,S\}$,
$$\overline{\text{MSE}}(\bm{w})=\sum_{s=1}^Sw_s^2V_s+B^2\left(\sum_{s=1}^S(p_s-w_s)\right)^2.$$
Therefore, $\bm{w}^M_B$ is the minimizer of
$$\sum_{s=1}^Sw_s^2V_s+B^2\left(\sum_{s=1}^S(p_s-w_s)\right)^2,$$
subject to
$$w_s-p_s\leq 0 \text{ for all }s.$$
The objective function is convex, and the inequality constraints are continuously differentiable and concave. Therefore, the Karush-Kuhn-Tucker conditions for optimality are also sufficient.

\medskip
The Lagrangian of this problem is
$$L(\bm{w},\bm{\mu})=\sum_{s=1}^Sw_s^2V_s+B^2\left(\sum_{s=1}^S(p_s-w_s)\right)^2+\sum_{s=1}^S 2\mu_s(w_s-p_s).$$
The Karush-Kuhn-Tucker conditions for optimality are
\begin{align}
&w^M_{B,s}V_s-B^2\left(1-\sum_{s'=1}^S w^M_{B,s'}\right)+\mu_s=0 \nonumber\\
&w^M_{B,s}\leq p_s\nonumber\\
&\mu_s \geq 0\nonumber\\
&\mu_s(w^M_{B,s}-p_s)=0.\label{eq:KKT conditions}
\end{align}
Those conditions are equivalent to
\begin{align}
&w^M_{B,s}=\min\left(\frac{1}{V_s}B^2\left(1-\sum_{s'=1}^S w^M_{B,s'}\right),p_s\right)\label{eq:w_opt}\\
&\mu_s=\max\left(0,B^2\left(1-\sum_{s'=1}^S w^M_{B,s'}\right)-p_sV_s\right)\nonumber.
\end{align}
As
\begin{align*}
&\frac{1}{V_s}B^2\left(1-\sum_{s'=1}^S w^M_{B,s'}\right)< p_s\nonumber\\
\Leftrightarrow& B^2\left(1-\sum_{s'=1}^S w^M_{B,s'}\right)<p_sV_s,
\end{align*}
\eqref{eq:w_opt} implies that
\begin{align}\label{eq:w_opt4}
& w^M_{B,s} < p_s \Rightarrow  w^M_{B,s+1} < p_{s+1}.
\end{align}
Let
$s^M=\min \{s \in \{1,...,S\}: w^M_{B,s} < p_s\}$, with the convention that $s^M=S+1$ if the set is empty. It follows from Equations \eqref{eq:w_opt} and \eqref{eq:w_opt4} that
\begin{align}
&w^M_{B,s}=p_s \text{ for all } s<s^M \nonumber\\
&w^M_{B,s}=\frac{1}{V_s}B^2\left(1-\sum_{s'=1}^S w^M_{B,s'}\right) \text{ for all } s\geq s^M. \label{eq:w_opt3}
\end{align}
 \eqref{eq:w_opt3} implies that
\begin{equation}\label{eq:w_opt3.5}
\sum_{s=s^M}^S w^M_{B,s}=\frac{B^2\sum_{s=s^M}^S\frac{1}{V_s}}{1+B^2\sum_{s=s^M}^S\frac{1}{V_s}}\sum_{s=s^M}^Sp_s.
\end{equation}
Plugging this equation into  \eqref{eq:w_opt3} yields
\begin{align}
&w^M_{B,s}=p_s \text{ for all } s<s^M \nonumber\\
&w^M_{B,s}=\frac{1}{V_s}\frac{1}{\frac{1}{B^2}+\sum_{s'=s^M}^S\frac{1}{V_{s'}}}\sum_{s'=s^M}^Sp_{s'}  \text{ for all } s\geq s^M. \label{eq:w_opt5}
\end{align}
To conclude the proof, we have to show that $s^M=\underline{s}$. First,
\begin{align*}
&\overline{\text{MSE}}(\bm{p})-\overline{\text{MSE}}\left(p_1,...,p_{S-1},\frac{\frac{1}{V_S}}{\frac{1}{B^2}+\frac{1}{V_S}}p_S\right)\\
=&p_S^2V_S-\left(p_S^2\left(\frac{\frac{1}{V_S}}{\frac{1}{B^2}+\frac{1}{V_S}}\right)^2V_S+B^2p_S^2\left(\frac{\frac{1}{B^2}}{\frac{1}{V_S}+\frac{1}{B^2}}\right)^2\right)\\
=&\frac{p_S^2}{\left(\frac{1}{B^2}+\frac{1}{V_S}\right)^2}\left(\frac{V_S}{B^4}+\frac{1}{B^2}\right)>0.
\end{align*}
Therefore, $s^M\in \{1,...,S\}$.
Second, assume that $s^M<\underline{s}$. Then, it follows from the definition of $\underline{s}$ that $w^M_{B,s^M}\geq p_{s^M}$, which contradicts the definition of $s^M$. If $\underline{s}=S$, we have shown that $s^M\in \{1,...,S\}$ and $s^M\geq \underline{s}$, which implies that $s^M=\underline{s}$: this completes the proof. Finally, if $\underline{s}<S$, assume that $s^M>\underline{s}$. Then, let
$$\tilde{w}_{\underline{s}}:=\frac{1}{V_{\underline{s}}}\frac{1}{\frac{1}{B^2}+\sum_{s={\underline{s}}}^S\frac{1}{V_s}}\sum_{s={\underline{s}}}^Sp_s<p_{\underline{s}},$$
where the inequality follows from the definition of $\underline{s}$.
Then, for any $w_{\underline{s}}: \tilde{w}_{\underline{s}}\leq w_{\underline{s}}<p_{\underline{s}}$,
\begin{align*}
&\overline{\text{MSE}}\left(\bm{w}^M_B\right)-\overline{\text{MSE}}\left(w^M_{B,1},...,w^M_{B,\underline{s}-1},w_{\underline{s}},w^M_{B,\underline{s}+1},...,w^M_{B,S}\right)\\
=&\left(p_{\underline{s}}^2-w_{\underline{s}}^2\right)V_{\underline{s}}+B^2\left(\left(\sum_{s=s^M}^S(p_s-w^M_{B,s})\right)^2-\left(\sum_{s=s^M}^S(p_s-w^M_{B,s})+p_{\underline{s}}-w_{\underline{s}}\right)^2\right)\\
=&\left(p_{\underline{s}}-w_{\underline{s}}\right)\left(\left(p_{\underline{s}}+w_{\underline{s}}\right)V_{\underline{s}}-B^2\left(2\sum_{s=s^M}^S(p_s-w^M_{B,s})+p_{\underline{s}}-w_{\underline{s}}\right)\right)\\
=&\left(p_{\underline{s}}-w_{\underline{s}}\right)\left(\left(p_{\underline{s}}+w_{\underline{s}}\right)V_{\underline{s}}-2B^2\times \sum_{s=s^M}^Sp_s\times \left(1-\frac{B^2\sum_{s=s^M}^S\frac{1}{V_s}}{1+B^2\sum_{s=s^M}^S\frac{1}{V_s}}\right)-B^2(p_{\underline{s}}-w_{\underline{s}})\right)\\
=&\left(p_{\underline{s}}-w_{\underline{s}}\right)\left(\left(p_{\underline{s}}+w_{\underline{s}}\right)V_{\underline{s}}-2\frac{1}{\frac{1}{B^2}+\sum_{s=s^M}^S\frac{1}{V_s}}\sum_{s=s^M}^Sp_s-B^2(p_{\underline{s}}-w_{\underline{s}})\right)\\
\geq &\left(p_{\underline{s}}-w_{\underline{s}}\right)\left(\left(p_{\underline{s}}+\tilde{w}_{\underline{s}}\right)V_{\underline{s}}-2\frac{1}{\frac{1}{B^2}+\sum_{s=\underline{s}}^S\frac{1}{V_s}}\sum_{s=\underline{s}}^Sp_s-B^2(p_{\underline{s}}-w_{\underline{s}})\right)\\
=&\left(p_{\underline{s}}-w_{\underline{s}}\right)\left(p_{\underline{s}}V_{\underline{s}}-\frac{1}{\frac{1}{B^2}+\sum_{s=\underline{s}}^S\frac{1}{V_s}}\sum_{s=\underline{s}}^Sp_s-B^2(p_{\underline{s}}-w_{\underline{s}})\right).
\end{align*}
The inequality follows from the fact that $\tilde{w}_{\underline{s}}\leq w_{\underline{s}}<p_{\underline{s}}$, and from Lemma \ref{lem_decreasingthrehsold}. The last equality follows from the definition of $\tilde{w}_{\underline{s}}$. It follows from the definition of $\underline{s}$ that $$p_{\underline{s}}V_{\underline{s}}-\frac{1}{\frac{1}{B^2}+\sum_{s=\underline{s}}^S\frac{1}{V_s}}\sum_{s=\underline{s}}^Sp_s>0.$$
Then, as $B^2(p_{\underline{s}}-w_{\underline{s}})$ can be made arbitrarily small by letting $w_{\underline{s}}$ go to $p_{\underline{s}}$, there exists
$w_{\underline{s}}$ such that $\tilde{w}_{\underline{s}}\leq w_{\underline{s}}<p_{\underline{s}}$ and $\overline{\text{MSE}}\left(\bm{w}^M_B\right)-\overline{\text{MSE}}\left(w^M_{B,1},...,w^M_{B,\underline{s}-1},w_{\underline{s}},w^M_{B,\underline{s}+1},...,w^M_{B,S}\right)>0$, a contradiction.
This completes the proof.

\subsection*{Proof of Lemma \ref{lem_worst-case-MSE-sign}}

$$\sum_{s=1}^S(w_s-p_s)\tau_s=\sum_{s=1}^S\left((w_s-p_s)_++(w_s-p_s)_-\right)\tau_s=\sum_{s=1}^S(w_s-p_s)_+\tau_s+\sum_{s=1}^S(w_s-p_s)_-\tau_s.$$
Therefore, under Assumption \ref{hyp:CATE_hat_sign}
$$B\sum_{s=1}^S(w_s-p_s)_-\leq \sum_{s=1}^S(w_s-p_s)\tau_s\leq B\sum_{s=1}^S(w_s-p_s)_+.$$
Therefore,
\begin{align*}
E\left(\left(\widehat{\tau}(\bm{w})-\tau\right)^2\right)=&\sum_{s=1}^Sw^2_sV(\widehat{\tau}_s)+\left(\sum_{s=1}^S(w_s-p_s)\tau_s\right)^2\\
\leq &\sum_{s=1}^Sw_{s}^2V_{s}+B^2\max\left[\left(\sum_{s=1}^{S}(w_s-p_s)_+\right)^2,\left(\sum_{s=1}^{S}(w_s-p_s)_-\right)^2\right],
\end{align*}
where the inequality follows from the previous display.

\subsection{Proof of Corollary \ref{cor_heteroscedasticite}}

\textit{Proof of Point 1}

Let $h_s=\sigma^2_{1,s}/\sigma^2\geq 1$.
As $\widehat{\tau}\left(\bm{w}^{\text{hom}}_{B}\right)$ is linear-minimax under \eqref{eq:homoscedasticity},
\begin{equation}\label{eq:cor_hetero1}
\sigma^2B^2\left(\sum_{s=1}^S\left(p_s-w^{\text{hom}}_{B,s}\right)\right)^2\leq \sigma^2\sum_{s=1}^S((p_s)^2-\left(w^{\text{hom}}_{B,s}\right)^2)(v_{0,s}+v_{1,s}).
\end{equation}
As for all $s$, $v_{1,s}\geq 0$, $(p_s)^2-\left(w^{\text{hom}}_{B,s}\right)^2\geq 0$, and $h_s\geq 1$,
\begin{equation}\label{eq:cor_hetero2}
\sigma^2\sum_{s=1}^S((p_s)^2-\left(w^{\text{hom}}_{B,s}\right)^2)(v_{0,s}+v_{1,s})\leq \sigma^2\sum_{s=1}^S((p_s)^2-\left(w^{\text{hom}}_{B,s}\right)^2)(v_{0,s}+h_sv_{1,s}).
\end{equation}
Combining \eqref{eq:cor_hetero1} and \eqref{eq:cor_hetero2} and rearranging proves the result.

\medskip
\textit{Proof of Point 2}

\medskip
Under the assumptions of Point 2 of the corollary,
the worst-case MSEs of $\widehat{\tau}(\bm{w}^{\text{hom}}_{B})$ and $\widehat{\tau}(\bm{p})$ are respectively equal to
$$\sigma^2\left(\sum_{s=1}^S\left(w^{\text{hom}}_{B,s}\right)^2(v_{0,s}+hv_{1,s})+B^2\left(\sum_{s=1}^S(p_s-w^{\text{hom}}_{B,s})\right)^2\right)$$
and
$$\sigma^2\sum_{s=1}^S(p_s)^2(v_{0,s}+hv_{1,s}).$$
Taking the difference between the two preceding displays, setting that difference lower than $0$ and rearranging yields the result.

\subsection{Proof of Lemma \ref{lem_worst-case-MSE_boundhet}}


Let
\begin{align*}
MSE(\bm{w},\bm{\tau})=&E\left(\left(\widehat{\tau}(\bm{w})-\tau\right)^2\right)=\sum_{s=1}^Sw_{s}^2V_{s}+\left(\sum_{s=1}^S(w_{s}-1/S)\tau_{s}\right)^2.
\end{align*}
$MSE(\bm{w},\bm{\tau})=MSE(\bm{w},-\bm{\tau})$. Therefore, without loss of generality assume that $\tau>0$. Then, to prove the result, as the first term does not depend on $(\tau_{(1)},...,\tau_{(S)})$, we need to maximize $$\left(\sum_{s=1}^S(w_{s}-1/S)\tau_{s}\right)^2=\left(\sum_{s=1}^S(w_{(s)}-1/S)\tau_{(s)}\right)^2$$ with respect to $(\tau_{(1)},...,\tau_{(S)})$, given $(w_{(1)},...,w_{(S)})$ and $\tau$, under the following constraints:
\begin{align}
&1/S\sum_{s=1}^S\tau_{(s)}=\tau \label{eq:maxbias_constraint1}\\
&-(B-1) \tau \leq \tau_s \leq (B+1) \tau. \label{eq:maxbias_constraint2}
\end{align}
Let
\begin{align*}
\tau^+_{(s)}=&(B+1)\tau1\{s\leq S/2\}-(B-1)\tau1\{s> S/2\}\\
\tau^-_{(s)}=&(B+1)\tau1\{s> S/2\}-(B-1)\tau1\{s\leq S/2\}.
\end{align*}
$(\tau^+_{(1)},...,\tau^+_{(S)})$ and $(\tau^-_{(1)},...,\tau^-_{(S)})$ verify \eqref{eq:maxbias_constraint1} and \eqref{eq:maxbias_constraint2}.
Let $(\tau_{(1)},...,\tau_{(S)})$  verify \eqref{eq:maxbias_constraint1} and \eqref{eq:maxbias_constraint2}.
Then, for any $s\leq S/2$,
\begin{align*}
&\tau_{(s)}-\tau^-_{(s)}\geq 0\\
& w_{(s)}-1/S\geq w_{(S/2)}-1/S.
\end{align*}
Therefore,
\begin{align}
\sum_{s=1}^{S/2}(w_{(s)}-1/S)(\tau_{(s)}-\tau^-_{(s)})\geq \left(w_{(S/2)}-1/S\right)\sum_{s=1}^{S/2}(\tau_{(s)}-\tau^-_{(s)}).\label{eq:maxbias_ineq1}
\end{align}
Similarly, for any $s> S/2$,
\begin{align*}
&\tau_{(s)}-\tau^-_{(s)}\leq 0\\
& w_{(s)}-1/S\leq w_{(S/2)}-1/S.
\end{align*}
Therefore,
\begin{align}
\sum_{s=S/2+1}^{S}(w_{(s)}-1/S)(\tau_{(s)}-\tau^-_{(s)})\geq \left(w_{(S/2)}-1/S\right)\sum_{s=S/2+1}^{S}(\tau_{(s)}-\tau^-_{(s)}).\label{eq:maxbias_ineq2}
\end{align}
Summing \eqref{eq:maxbias_ineq1} and \eqref{eq:maxbias_ineq2},
\begin{align}
\sum_{s=1}^{S}(w_{(s)}-1/S)(\tau_{(s)}-\tau^-_{(s)})\geq \left(w_{(S/2)}-1/S\right)\sum_{s=1}^{S}(\tau_{(s)}-\tau^-_{(s)})=0,\label{eq:maxbias_ineq3}
\end{align}
where the equality follows from the fact $\tau_{s}$ and $\tau^-_{s}$ satisfy \eqref{eq:maxbias_constraint1}. Using similar steps, one can show that
\begin{align}
\sum_{s=1}^{S}(w_{(s)}-1/S)(\tau^+_{(s)}-\tau_{(s)})\geq \left(w_{(S/2)}-1/S\right)\sum_{s=1}^{S}(\tau^+_{(s)}-\tau_{(s)})=0.\label{eq:maxbias_ineq6}
\end{align}
Combining \eqref{eq:maxbias_ineq3} and \eqref{eq:maxbias_ineq6},
\begin{align}\label{eq:maxbias_ineq7}
&\sum_{s=1}^{S}(w_{(s)}-1/S)\tau^-_{(s)} \leq \sum_{s=1}^{S}(w_{(s)}-1/S)\tau_{(s)}\leq \sum_{s=1}^{S}(w_{(s)}-1/S)\tau^+_{(s)}.
\end{align}
Therefore,
\begin{align*}
\left(\sum_{s=1}^{S}(w_{(s)}-1/S)\tau_{(s)}\right)^2 \leq \max \left[\left(\sum_{s=1}^{S}(w_{(s)}-1/S)\tau^-_{(s)}\right)^2,\left(\sum_{s=1}^{S}(w_{(s)}-1/S)\tau^+_{(s)}\right)^2\right].
\end{align*}
This proves the validity and the sharpness of the upper bound.

\subsection{Statement and proof of Lemma \ref{lem_getting_rid_max}}

Some results below rely on the following lemma.
\begin{lem}\label{lem_getting_rid_max}
\begin{enumerate}
  \item For any $\bm{w}\in \mathbb{R}^S_+$, if
  \begin{align*}
&\left|(B+1)\sum_{s=1}^{S/2}(w_{(s)}-1/S)-(B-1)\sum_{s=S/2+1}^{S}(w_{(s)}-1/S)\right|\\
> & \left|(B+1)\sum_{s=S/2+1}^{S}(w_{(s)}-1/S)-(B-1)\sum_{s=1}^{S/2}(w_{(s)}-1/S)\right|,
\end{align*}
then $(B+1)\sum_{s=1}^{S/2}(w_{(s)}-1/S)-(B-1)\sum_{s=S/2+1}^{S}(w_{(s)}-1/S)> 0$.
  \item If $B>0$, for any $\bm{w}\in \mathbb{R}^S_+$ such that $s\mapsto w_{(s)}$ is not constant,
\begin{align*}
&\sum_{s=1}^{S}w_{(s)}>1\\
\Leftrightarrow & \left|(B+1)\sum_{s=1}^{S/2}(w_{(s)}-1/S)-(B-1)\sum_{s=S/2+1}^{S}(w_{(s)}-1/S)\right|\\
> & \left|(B+1)\sum_{s=S/2+1}^{S}(w_{(s)}-1/S)-(B-1)\sum_{s=1}^{S/2}(w_{(s)}-1/S)\right|.
\end{align*}
\end{enumerate}
\end{lem}

\medskip
\textbf{Proof of Lemma \ref{lem_getting_rid_max}}

\medskip
\textbf{Proof of Point 1}

\medskip
In view of \eqref{eq:ineq_max}, if $(B+1)\sum_{s=1}^{S/2}(w_{(s)}-1/S)-(B-1)\sum_{s=S/2+1}^{S}(w_{(s)}-1/S)\leq 0$,
\begin{align*}
&\left|(B+1)\sum_{s=1}^{S/2}(w_{(s)}-1/S)-(B-1)\sum_{s=S/2+1}^{S}(w_{(s)}-1/S)\right|\\
\leq & \left|(B+1)\sum_{s=S/2+1}^{S}(w_{(s)}-1/S)-(B-1)\sum_{s=1}^{S/2}(w_{(s)}-1/S)\right|.
\end{align*}
The result follows by contraposition.

\medskip
\textbf{Proof of Point 2}

\medskip
The inequality in \eqref{eq:ineq_max} is actually strict if and only if
$$B\left(\sum_{s=1}^{S/2}w_{(s)}-\sum_{s=S/2+1}^{S}w_{(s)}\right)>0,$$
which holds because we have assumed that $B>0$ and $s\mapsto w_{(s)}$ is not constant.
Therefore,
\footnotesize
\begin{align*}
&\left|(B+1)\sum_{s=1}^{S/2}(w_{(s)}-1/S)-(B-1)\sum_{s=S/2+1}^{S}(w_{(s)}-1/S)\right|> \left|(B+1)\sum_{s=S/2+1}^{S}(w_{(s)}-1/S)-(B-1)\sum_{s=1}^{S/2}(w_{(s)}-1/S)\right|\\
\Leftrightarrow& (B+1)\sum_{s=1}^{S/2}(w_{(s)}-1/S)-(B-1)\sum_{s=S/2+1}^{S}(w_{(s)}-1/S)>-(B+1)\sum_{s=S/2+1}^{S}(w_{(s)}-1/S)+(B-1)\sum_{s=1}^{S/2}(w_{(s)}-1/S)\\
\Leftrightarrow& \sum_{s=1}^{S}w_{(s)}>1.
\end{align*}
\normalsize
The second equivalence follows after some algebra. The first equivalence is due to the fact that for any real numbers $a$ and $b$, if $a>b$, then $|a|> |b|\Leftrightarrow a > -b$.
First, assume that $a\leq -b$. Then, as by assumption $a>b$, $b<a\leq-b$. Therefore, $a\leq-b$, $-a<-b$, $|b|=-b$, so $|a|\leq |b|$. Thus, $a\leq -b \Rightarrow |a|\leq |b|$. By contraposition,
$|a|> |b|\Rightarrow a > -b$. Second, assume that $a > -b$. As by assumption $a>b$, then $a>|b|$ and $|a|=a$, hence $|a|>|b|$. Therefore, $a> -b \Rightarrow |a|>|b|$.

\subsection{Proof of Lemma \ref{lem_average_opt_weights_lowerthanone}}

\textbf{Proof of Point 1}

\medskip
\textbf{Case 1: $s\mapsto w^H_{[s]}$ is not constant}

\medskip
Assume that $s\mapsto w^H_{[s]}$ is not constant, and $\sum_{s=1}^{S}w^H_{[s]}>1.$ We cannot have $w^H_{[s]}=0$ for all $s$, and let $s_0:\max\{s:w^H_{[s]}>0\}$.
Then, for a strictly positive $$\eps<\min\left(\sum_{s=1}^{S}w^H_{[s]}-1,w^H_{[s_0]}S\right),$$ let $\tilde{w}_{[s]}=w^H_{[s]}-\eps/S$ if $s\leq s_0$, $\tilde{w}_{[s]}=w^H_{[s]}=0$ otherwise. $\tilde{w}_{[s_0]}=w^H_{[s_0]}-\eps/S>0$. Therefore, $\bm{\tilde{w}}\in \mathbb{R}^S_+$ and $s\mapsto \tilde{w}_{[s]}$ is decreasing. Then, $$\sum_{s=1}^{S}\tilde{w}_{[s]}=\sum_{s=1}^{S}w^H_{[s]}-\eps \times s_0/S\geq \sum_{s=1}^{S}w^H_{[s]}-\eps>1,$$
and we also have that $s\mapsto \tilde{w}_{(s)}$ is not constant.

\medskip
\textbf{Subcase 1: $B>0$}

\medskip
Then, it follows from the first and second points of Lemma \ref{lem_getting_rid_max} that
$$\overline{SQB}^H(\bm{\tilde{w}})=\left((B+1)\sum_{s=1}^{S/2}(\tilde{w}_{[s]}-1/S)-(B-1)\sum_{s=S/2+1}^{S}(\tilde{w}_{[s]}-1/S)\right)^2,$$
and
$$(B+1)\sum_{s=1}^{S/2}(\tilde{w}_{[s]}-1/S)-(B-1)\sum_{s=S/2+1}^{S}(\tilde{w}_{[s]}-1/S)>0.$$
As $\sum_{s=1}^{S}w^H_{[s]}>1$ and $s\mapsto w^H_{[s]}$ is not constant, it follows from the second point of Lemma \ref{lem_getting_rid_max} that
$$\overline{SQB}^H(\bm{w}^H)=\left((B+1)\sum_{s=1}^{S/2}(w^H_{[s]}-1/S)-(B-1)\sum_{s=S/2+1}^{S}(w^H_{[s]}-1/S)\right)^2.$$
Assume $s_0>S/2$. Then,
\begin{align*}
&(B+1)\sum_{s=1}^{S/2}(w^H_{[s]}-1/S)-(B-1)\sum_{s=S/2+1}^{S}(w^H_{[s]}-1/S)\\
-&\left((B+1)\sum_{s=1}^{S/2}(\tilde{w}_{[s]}-1/S)-(B-1)\sum_{s=S/2+1}^{S}(\tilde{w}_{[s]}-1/S)\right)\\
=&(B+1)\sum_{s=1}^{S/2}w^H_{[s]}-(B-1)\sum_{s=S/2+1}^{S}w^H_{[s]}-\left((B+1)\sum_{s=1}^{S/2}\tilde{w}_{[s]}-(B-1)\sum_{s=S/2+1}^{S}\tilde{w}_{[s]}\right)\\
=&(B+1)\sum_{s=1}^{S/2}(w^H_{[s]}-\tilde{w}_{[s]})-(B-1)\sum_{s=S/2+1}^{s_0}(w^H_{[s]}-\tilde{w}_{[s]})\\
=&\eps\left((B+1)/2-(B-1)(s_0-S/2)/S\right)>0.
\end{align*}
Similarly, if $s_0\leq S/2$,
\begin{align*}
&(B+1)\sum_{s=1}^{S/2}(w^H_{[s]}-1/S)-(B-1)\sum_{s=S/2+1}^{S}(w^H_{[s]}-1/S)\\
-&\left((B+1)\sum_{s=1}^{S/2}(\tilde{w}_{[s]}-1/S)-(B-1)\sum_{s=S/2+1}^{S}(\tilde{w}_{[s]}-1/S)\right)\\
=&\eps (B+1)s_0/S>0.
\end{align*}
Therefore, we always have
$\overline{SQB}^H(\bm{w}^H)>\overline{SQB}^H(\bm{\tilde{w}})$. Then, as $0\leq \tilde{w}_s\leq w^H_s$ for all $s$ and the second inequality is strict for some $s$, $V(\tilde{\bm{w}})<V(\bm{w}^H)$. Therefore, $\overline{\text{MSE}}^H(\bm{w}^H)>\overline{\text{MSE}}^H(\bm{\tilde{w}})$, a contradiction.

\medskip
\textbf{Subcase 2: $B=0$}

\medskip
Then,
$$\overline{SQB}^H(\bm{\tilde{w}})=\left(\sum_{s=1}^{S}\tilde{w}_{(s)}-1\right)^2<\left(\sum_{s=1}^{S}w^H_{[s]}-1\right)^2=\overline{SQB}^H(\bm{w}^H).$$
As before, $V(\tilde{\bm{w}})<V(\bm{w}^H)$. Therefore, $\overline{\text{MSE}}^H(\bm{w}^H)>\overline{\text{MSE}}^H(\bm{\tilde{w}})$, a contradiction.

\medskip
\textbf{Case 2: $s\mapsto w^H_{[s]}$ constant}

\medskip
Assume that there is a number $k>1/S$ such that $w^H_{[s]}=k$ for all $s$. Then, let $\tilde{w}_{s}=1/S$.
$\overline{SQB}^H(\bm{\tilde{w}})=0\leq \overline{SQB}^H(\bm{w}^H)$.
Then, as $0< \tilde{w}_s<w^H_s$ for all $s$, $V(\tilde{\bm{w}})<V(\bm{w}^H)$. Therefore, $\overline{\text{MSE}}^H(\bm{w}^H)>\overline{\text{MSE}}^H(\bm{\tilde{w}})$, a contradiction.

\medskip
\textbf{Proof of Point 2}

\medskip
\eqref{eq:maxbias_ineq7} implies that for any $\bm{w}\in \mathbb{R}^S_+$,
\begin{align}\label{eq:ineq_max}
&(B+1)\sum_{s=1}^{S/2}(w_{(s)}-1/S)-(B-1)\sum_{s=S/2+1}^{S}(w_{(s)}-1/S)\nonumber\\
\geq& (B+1)\sum_{s=S/2+1}^{S}(w_{(s)}-1/S)-(B-1)\sum_{s=1}^{S/2}(w_{(s)}-1/S).
\end{align}
Then,
\begin{align*}
& (B+1)\sum_{s=1}^{S/2}(w_{(s)}-1/S)-(B-1)\sum_{s=S/2+1}^{S}(w_{(s)}-1/S)\\
+& \left((B+1)\sum_{s=S/2+1}^{S}(w_{(s)}-1/S)-(B-1)\sum_{s=1}^{S/2}(w_{(s)}-1/S)\right)\\
=& 2\sum_{s=1}^{S}w_{(s)}-2.
\end{align*}
The previous display and \eqref{eq:ineq_max} imply that if $\sum_{s=1}^{S}w_{(s)}\leq 1$,
\begin{align*}
&-(B+1)\sum_{s=S/2+1}^{S}(w_{(s)}-1/S)+(B-1)\sum_{s=1}^{S/2}(w_{(s)}-1/S)\nonumber\\
\geq&(B+1)\sum_{s=1}^{S/2}(w_{(s)}-1/S)-(B-1)\sum_{s=S/2+1}^{S}(w_{(s)}-1/S)\nonumber\\
\geq& (B+1)\sum_{s=S/2+1}^{S}(w_{(s)}-1/S)-(B-1)\sum_{s=1}^{S/2}(w_{(s)}-1/S),
\end{align*}
hence the result.

\subsection{Proof of Lemma \ref{lem_permutation}}

Assume that there exists $s_0$ such that $w^H_{s_0}<w^H_{s_0+1}$. Then, let $\tilde{w}_{s_0}=w^H_{s_0+1}$, $\tilde{w}_{s_0+1}=w^H_{s_0}$, and $\tilde{w}_{s}=w^H_{s}$ for all $s\notin \{s_0,s_0+1\}$.
Then,
\begin{align*}
V(\bm{w}^H)-V(\bm{\tilde{w}})=&\left((w^H_{s_0})^2-(w^H_{s_0+1})^2\right)\left(V_{s_0}-V_{s_0+1}\right)> 0.
\end{align*}
As $\tilde{w}_{s}=w^H_{\sigma(s)}$ for a permutation $\sigma()$,
$s\mapsto \tilde{w}_{[s]}$ is decreasing. Therefore, $\overline{SQB}^H(\bm{w}^H)=\overline{SQB}^H(\bm{\tilde{w}})$.
Then, $\overline{\text{MSE}}^H(\tilde{\bm{w}})<\overline{\text{MSE}}^H(\bm{w}^H)$, a contradiction.


\subsection{Proof of Theorem \ref{thm_minimax2}}

\textbf{Proof of Point 1}

\medskip
It follows from Point 1 of Lemma \ref{lem_average_opt_weights_lowerthanone} and Lemma \ref{lem_permutation} that $\bm{w}^H \in \mathcal{R}$. Then,
$\bm{w}^H=\argmin_{\bm{w}\in \mathcal{R}} \overline{\text{MSE}}^H(\bm{w}).$
Then, it follows from Lemma \ref{lem_permutation} that
\begin{align*}
\overline{SQB}^H(\bm{w})=&\tau^2\max\left[\left((B+1)\sum_{s=1}^{S/2}(w_{s}-1/S)-(B-1)\sum_{s=S/2+1}^{S}(w_{s}-1/S)\right)^2,\right.\\
&\left.\left((B+1)\sum_{s=S/2+1}^{S}(w_{s}-1/S)-(B-1)\sum_{s=1}^{S/2}(w_{s}-1/S)\right)^2\right].
\end{align*}
Then, for any $\bm{w}\in \mathcal{R}$,
\begin{align*}
&\max\left[\left((B+1)\sum_{s=1}^{S/2}(w_{s}-1/S)-(B-1)\sum_{s=S/2+1}^{S}(w_{s}-1/S)\right)^2,\right.\\
&\left.\left((B+1)\sum_{s=S/2+1}^{S}(w_{s}-1/S)-(B-1)\sum_{s=1}^{S/2}(w_{s}-1/S)\right)^2\right]\\
=&\left((B+1)\sum_{s=S/2+1}^{S}(w_{s}-1/S)-(B-1)\sum_{s=1}^{S/2}(w_{s}-1/S)\right)^2\\
=&\left((B-1)\sum_{s=1}^{S/2}w_{s}-(B+1)\sum_{s=S/2+1}^{S}w_{s}+1\right)^2,
\end{align*}
where the first equality follows from Point 2 of Lemma \ref{lem_average_opt_weights_lowerthanone}, and the second follows after some algebra. The two preceding displays imply that for any $\bm{w}\in \mathcal{R}$, $\overline{\text{MSE}}^H(\bm{w})=\overline{\text{MSE}}^{H,d}(\bm{w}).$

\medskip
\textbf{Proof of Point 2}

\medskip
If $w_{1}^H=0$, then $w_{s}^H=0$ for all $s$ and the result trivially holds. Henceforth, we assume that $w_{1}^H>0$.

\medskip
$\mathcal{R}$ is a convex subset of $\mathbb{R}^S$, $\overline{\text{MSE}}^{H,d}(\bm{w})$ is strictly convex and continuously differentiable on $\mathcal{R}$, the inequality constraints are linear, and Slater's condition holds (for instance, $(1/S,...,1/S)\in \mathcal{R}$, and it weakly satisfies all inequality constraints, which is sufficient for Slater's condition to hold as the inequality constraints are linear).  Therefore, $\bm{w}^H$ is the solution to a minimization problem whose Karush-Kuhn-Tucker conditions are necessary and sufficient for optimality.

\medskip
The Lagrangian of the minimization problem is
\begin{align*}
L(\bm{w},\bm{\mu},\lambda)=&\sum_{s=1}^{S}w_{s}^2V_{s}+\tau^2\left((B-1)\sum_{s=1}^{S/2}w_{s}-(B+1)\sum_{s=S/2+1}^{S}w_{s}+1\right)^2
\\
+&\sum_{s=1}^{S-1} 2\mu_s(w_{s+1}-w_s)-2\mu_Sw_{S}+2\lambda \left(\sum_{s=1}^{S}w_{s}-1\right).
\end{align*}
The Karush-Kuhn-Tucker conditions for optimality are:
\begin{align}
&w^H_1 V_{1}+(B-1)\tau^2\left((B-1)\sum_{s=1}^{S/2}w^H_{s}-(B+1)\sum_{s=S/2+1}^{S}w^H_{s}+1\right)-\mu_{1}+\lambda=0\nonumber\\
&\forall s \in \{2,...,S/2\}:w^H_s V_{s}+(B-1)\tau^2\left((B-1)\sum_{s=1}^{S/2}w^H_{s}-(B+1)\sum_{s=S/2+1}^{S}w^H_{s}+1\right)+\mu_{s-1}-\mu_{s}+\lambda=0\nonumber\\
&\forall s \in \{S/2+1,...,S\}:w^H_s V_{s}-(B+1)\tau^2\left((B-1)\sum_{s=1}^{S/2}w^H_{s}-(B+1)\sum_{s=S/2+1}^{S}w^H_{s}+1\right)+\mu_{s-1}-\mu_{s}+\lambda=0\nonumber\\
&\forall s \in \{1,...,S-1\}:~w^H_{s+1}-w^H_s \leq 0\nonumber\\
&-w^H_S \leq 0\nonumber\\
&\forall s \in \{1,...,S\}:~~\mu_s \geq 0\nonumber\\
&\forall s \in \{1,...,S-1\}:~~\mu_s(w^H_{s+1}-w^H_s)=0\nonumber\\
&\mu_Sw^H_{S}=0\nonumber\\
&\sum_{s=1}^{S}w^H_{s}-1\leq 0\nonumber\\
&\lambda \geq 0\nonumber\\
&\lambda \left(\sum_{s=1}^{S}w^H_{s}-1\right)=0.\label{eq:KKT conditions2}
\end{align}

\medskip
First, we show by contradiction that $w^H_{S/2}=w^H_{1}$. If $w^H_{S/2}<w^H_{1}$, let $s_1=\min \{s\in \{2,...,S/2\}:w^H_{s}<w^H_{1}\}$.
For any strictly positive 
$\eps<\min\left(w^H_{1}-w^H_{s_1},w^H_{1}\right),$
let $\tilde{w}_{s}=w^H_{s}-\eps$ for $s\in \{1,...,s_1-1\}$, $\tilde{w}_{s}=w^H_{s}$ otherwise. $\bm{\tilde{w}}\in\mathbb{R}^S_+$, $s\mapsto \tilde{w}_{s}$ is decreasing, and
$$\sum_{s=1}^{S}\tilde{w}_{s}\leq \sum_{s=1}^{S}w^H_{s}\leq 1.$$
Then, as $B\geq 1$, $(B-1)\sum_{s=1}^{S}\tilde{w}_{s}\geq 0$. Moreover, as $\sum_{s=1}^{S}\tilde{w}_{s}\leq 1$ and $s\mapsto \tilde{w}_{s}$ decreasing, $\sum_{s=S/2+1}^{S}\tilde{w}_{s}\leq 1/2$. Therefore,
\begin{equation}\label{eq:negativebias}
(B-1)\sum_{s=1}^{S/2}\tilde{w}_{s}-(B+1)\sum_{s=S/2+1}^{S}\tilde{w}_{s}+1=(B-1)\sum_{s=1}^{S}\tilde{w}_{s}-2\sum_{s=S/2+1}^{S}\tilde{w}_{s}+1\geq 0.
\end{equation}
Moreover, as $(B-1)\geq 0$ and
$\sum_{s=1}^{s_1-1}\tilde{w}_{s}<\sum_{s=1}^{s_1-1}w^H_{s},$
$$(B-1)\sum_{s=1}^{S/2}\tilde{w}_{s}-(B+1)\sum_{s=S/2+1}^{S}\tilde{w}_{s}+1 \leq (B-1)\sum_{s=1}^{S/2}w^H_{s}-(B+1)\sum_{s=S/2+1}^{S}w^H_{s}+1.$$
The two preceding displays imply that $\overline{SQB}^H(\tilde{\bm{w}})\leq\overline{SQB}^H(\bm{w}^H)$. Moreover, as $0\leq \tilde{w}_{s}\leq w^H_{s}$ for all $s$ with a strict inequality for some $s$, $V(\tilde{\bm{w}})<V(\bm{w}^H)$. Then, $\overline{\text{MSE}}^H(\tilde{\bm{w}})<\overline{\text{MSE}}^H(\bm{w}^H)$, a contradiction.

\medskip
Next, we show by induction that for every $k\in \{S/2+1,...,s_0\}$, $w_{k}^H=w_{1}^H.$
First, summing the FOC conditions attached to $w^H_1$, $w^H_2$,..., $w_{S/2}^H$ yields:
\begin{equation*}\label{eq:sum_FOCs1}
\mu_{S/2}=w^H_1 \sum_{s=1}^{S/2}V_{s}+S/2(B-1)\tau^2\left((B-1)\sum_{s=1}^{S/2}w^H_{s}-(B+1)\sum_{s=S/2+1}^{S}w^H_{s}+1\right)+S/2\lambda.
\end{equation*}
Using the same steps as those used to show \eqref{eq:negativebias}, one can show that
\begin{equation}\label{eq:negativebias2}
\left((B-1)\sum_{s=1}^{S/2}w^H_{s}-(B+1)\sum_{s=S/2+1}^{S}w^H_{s}+1\right)\geq 0.
\end{equation}
Then, as $S/2(B-1)\tau^2\geq 0$, $S/2\lambda\geq 0$, and $w^H_1 \sum_{s=1}^{S/2}V_{s}>0$, $\mu_{S/2}>0$.
Therefore, $w_{S/2+1}^H=w_{1}^H$. Then, assume that for some $k\in \{S/2+1,...,\lfloor S(1-1/(B+1)) \rfloor\}$, $w_{j}^H=w_{1}^H$ for all $j\in \{S/2+1,...,k\}$. Summing the FOC conditions attached to $w_{1}^H$, ..., $w_{k}^H$,
\begin{align*}
\mu_{k}=&w^H_1 \sum_{s=1}^{k}V_{s}+(S/2(B-1)-(k-S/2)(B+1))\tau^2\left((B-1)\sum_{s=1}^{S/2}w^H_{s}-(B+1)\sum_{s=S/2+1}^{S}w^H_{s}+1\right)+k\lambda.
\end{align*}
As $k\leq \lfloor S(1-1/(B+1)) \rfloor$, $S/2(B-1)-(k-S/2)(B+1)\geq 0$. Then,
$\mu_{k}>0$ by the same arguments as before. Therefore, $w_{1}^H=w_{k+1}^H$. This proves the result.

\medskip
\textbf{Proof of Point 3}

\medskip
For all $s$, let $w^H_{s}=\frac{\tau^2}{\frac{1}{S^2}\sum_{s'=1}^{S}V_{s'}+\tau^2}1/S$. Let $\lambda=0$.
For all $s\leq S/2$ let
\begin{align*}
\mu_{s}=&\frac{\tau^2}{\frac{1}{S^2}\sum_{s'=1}^{S}V_{s'}+\tau^2}  \frac{1}{S}\sum_{s'=1}^{s}V_{s'}+s(B-1)\tau^2\frac{\frac{1}{S^2}\sum_{s'=1}^{S}V_{s'}}{\frac{1}{S^2}\sum_{s'=1}^{S}V_{s'}+\tau^2},
\end{align*}
and for all $s\geq S/2+1$ let
\begin{align*}
\mu_{s}=&\frac{\tau^2}{\frac{1}{S^2}\sum_{s'=1}^{S}V_{s'}+\tau^2}  \frac{1}{S}\sum_{s'=1}^{s}V_{s'}+(S/2(B-1)-(s-S/2)(B+1))\tau^2\frac{\frac{1}{S^2}\sum_{s'=1}^{S}V_{s'}}{\frac{1}{S^2}\sum_{s'=1}^{S}V_{s'}+\tau^2},
\end{align*}
thus ensuring that the FOC attached to $w^H_{s}$ holds for all $s$. For all $s\leq S-1$,  $w^H_{s+1}-w^H_{s}=0$ and $\mu_s(w^H_{s+1}-w^H_{s})=0$. $-w^H_{S}<0$, $\mu_S=0$, and $-w^H_{S}\mu_S=0$. For all $s\leq S/2$,
$\mu_{s}\geq 0.$ For all $s\geq S/2+1$,
$$\mu_{s}\geq 0 \Leftrightarrow \sum_{s'=1}^{s}V_{s'}/\sum_{s'=1}^{S}V_{s'}\geq s(B+1)/S-B.$$
By Lemma \ref{lem:positive_mus}, this condition holds.
Finally, $\sum_{s'=1}^{S}w^H_{s'}-1<0$, $\lambda=0$, and $\lambda \left(\sum_{s'=1}^{S}w^H_{s'}-1\right)=0$. Therefore, $(\bm{w}^H,\bm{\mu},\lambda)$ satisfies all the conditions for optimality in \eqref{eq:KKT conditions2}.

\subsection{Proof of Proposition \ref{prop:firststepFLCI}}

The Lagrangian of this problem is
$$L(\bm{w},\bm{\mu},\bm{\lambda})=\sum_{s=1}^Sw_s^2V_s+2\lambda \left(B\sum_{s=1}^S(p_s-w_s)-M\right)+\sum_{s=1}^S 2\mu_s(w_s-p_s).$$
The Karush-Kuhn-Tucker conditions for optimality are
\begin{align}
&w^{\text{CI}}_{M,s}V_s-\lambda B+\mu_s=0 \nonumber\\
& B\sum_{s=1}^S(p_s-w^{\text{CI}}_{M,s})\leq M \nonumber\\
& \lambda \geq 0 \nonumber\\
& \lambda\left(B\sum_{s=1}^S(p_s-w^{\text{CI}}_{M,s})-M\right)=0 \nonumber\\
&w^{\text{CI}}_{M,s}\leq p_s\nonumber\\
&\mu_s \geq 0\nonumber\\
&\mu_s(w^{\text{CI}}_{M,s}-p_s)=0.\label{eq:KKT conditions_CIFL}
\end{align}
\eqref{eq:KKT conditions_CIFL} implies that
\begin{equation}\label{eq::AK_Min1_step0}
w^{\text{CI}}_{M,s}=\min(p_s,\lambda B/V_s)
\end{equation}
and
\begin{equation}\label{eq::AK_Min1_step0'}
\mu_s=\max(\lambda B-p_sV_s,0).
\end{equation}
One cannot have $\lambda=0$, as this would imply $\mu_s=w^{\text{CI}}_{M,s}=0$ for all $s$, but then we would have $B\sum_{s=1}^S(p_s-w^{\text{CI}}_{M,s})> M$. Therefore, $\lambda>0$, and
\begin{equation}\label{eq::AK_Min1_step1}
B\sum_{s=1}^S(p_s-w^{\text{CI}}_{M,s})=M.
\end{equation}
Let $s_{\text{ci}}=\min\{s\in \{1,...,S\}:w^{\text{CI}}_{M,s}<p_s\}$. The set cannot be empty (if $w^{\text{CI}}_{M,S}=p_S$, then $B\sum_{s=1}^S(p_s-w^{\text{CI}}_{M,s})=0<M$) so $s_{\text{ci}}$ is well defined.
Then, as $\lambda B/V_s<p_s\Rightarrow \lambda B/V_{s+1}<p_{s+1}$,
\begin{equation}\label{eq::AK_Min1_step2}
w^{\text{CI}}_{M,s}=p_s1\{s<s_{\text{ci}}\}+\lambda B/V_s1\{s\geq s_{\text{ci}}\}.
\end{equation}
Plugging \eqref{eq::AK_Min1_step2} into \eqref{eq::AK_Min1_step1} and solving for $\lambda$,
\begin{equation}\label{eq::AK_Min1_step3}
\lambda=\frac{\sum_{s=s_{\text{ci}}}^Sp_s-M/B}{B\sum_{s=s_{\text{ci}}}^S1/V_s}=\lambda(s_{\text{ci}}).
\end{equation}
In view of all the above, $\lambda(s_{\text{ci}})>0$, $\lambda(s_{\text{ci}})B/V_{s_{\text{ci}}}<p_{s_{\text{ci}}}$, and $\forall s'<s:\lambda(s_{\text{ci}})B/V_{s'}\geq p_{s'}$. Therefore, $s_{\text{ci}}\in \mathcal{S}_{\text{ci}}$.

\subsection{Proof of Theorem \ref{thm:CB}}

$\mathcal{R}^{\text{CB}}$ is convex and the objective function is strictly convex and differentiable on $\mathcal{R}^{\text{CB}}$ because $\sum_{s=1}^Sw_s^2V_s>0$.
The Lagrangian of this problem is
$$L(\bm{w}_{-1},\bm{\mu})=z_{1-\alpha}\sqrt{\sum_{s=1}^Sw_s^2V_s}+B\sum_{s=1}^S(p_s-w_s)+\sum_{s=1}^S \mu_s(w_s-p_s).$$
The Karush-Kuhn-Tucker conditions for optimality are
\begin{align}
&w^{\text{CB}}_sz_{1-\alpha}V_s/\sigma(\bm{w}^{\text{CB}})-B+\mu_s=0 \nonumber\\
&w^{\text{CB}}_s\leq p_s\nonumber\\
&\mu_s \geq 0\nonumber\\
&\mu_s(w^{\text{CB}}_s-p_s)=0.\label{eq:KKT conditions3}
\end{align}
\eqref{eq:KKT conditions3} implies that
\begin{equation}\label{eq:CB_step0}
w^{\text{CB}}_{s}=\min(p_s,\sigma(\bm{w}^{\text{CB}}) B/(z_{1-\alpha}V_s))
\end{equation}
and
\begin{equation}\label{eq:CB_step0'}
\mu_s=\max(B-p_sz_{1-\alpha}V_s/\sigma(\bm{w}^{\text{CB}}),0).
\end{equation}
If $p_SV_S\leq \sigma(\bm{p}) B/z_{1-\alpha}$, $\bm{p}$ and the corresponding $\bm{\mu}$ satisfy \eqref{eq:KKT conditions3}, so $\bm{w}^{\text{CB}}=\bm{p}$. On the other hand, if $p_SV_S>\sigma(\bm{p}) B/z_{1-\alpha}$, $\bm{p}$ cannot satisfy \eqref{eq:KKT conditions3} (this would imply $\mu_S<0$). Then, let $s_{\text{cb}}=\min\{s\in \{1,...,S\}:w^{\text{CB}}_{s}<p_s\}$. As $\sigma(\bm{w}^{\text{CB}}) B/(z_{1-\alpha}V_s)<p_s\Rightarrow \sigma(\bm{w}^{\text{CB}}) B/(z_{1-\alpha}V_{s+1})<p_{s+1}$,
\begin{equation}\label{eq:CB_step2}
w^{\text{CB}}_{s}=p_s1\{s<s_{\text{cb}}\}+\sigma(\bm{w}^{\text{CB}}) B/(z_{1-\alpha}V_s)1\{s\geq s_{\text{cb}}\}.
\end{equation}
Plugging \eqref{eq:CB_step2} into $$\sigma^2(\bm{w}^{\text{CB}})=\sum_{s=1}^S(w^{\text{CB}}_s)^2V_s$$
and solving for $\sigma^2(\bm{w}^{\text{CB}})$,
\begin{equation}\label{eq:CB_step3}
\sigma^2(\bm{w}^{\text{CB}})=\frac{\sum_{s=1}^{s_{\text{cb}}-1}p^2_sV_s}{1-\frac{B^2}{z^2_{1-\alpha}}\sum_{s=s_{\text{cb}}}^{S}1/V_s}=\sigma^2(s_{\text{cb}}).
\end{equation}
In view of all the above, $\sigma^2(s_{\text{cb}})>0$, $\sigma^2(s_{\text{cb}}) B/(z_{1-\alpha}V_{s_{\text{cb}}})<p_{s_{\text{cb}}}$, and $\forall s'<s:\sigma^2(s_{\text{cb}}) B/(z_{1-\alpha}V_{s'})\geq p_{s'}$. Therefore, $s_{\text{cb}}\in \mathcal{S}_{\text{cb}}$.

\subsection{Proof of Proposition \ref{prop:conditions_under_whichpowerlower}}

For any $\lambda \in \left[1,\underset{s\in \{1,...,S\}}{\min}p_s/w_s\right]$, let
$$g:\lambda \mapsto \frac{\overline{|B|}(\lambda\bm{w})}{\sigma(\lambda\bm{w})}=\frac{B\sum_{s=1}^S(p_s-\lambda w_s)}{\lambda\sigma(\bm{w})}=B\frac{1-\lambda \sum_{s=1}^S w_s}{\lambda\sigma(\bm{w})},$$
where the first equality follows from the fact that, as $\lambda \leq \underset{s\in \{1,...,S\}}{\min}p_s/w_s$, $\lambda w_s\leq p_s ~\forall s$.
$$\frac{\partial g}{\partial \lambda}=B\frac{-\sum_{s=1}^S w_s \lambda\sigma(\bm{w})-\sigma(\bm{w})(1-\lambda \sum_{s=1}^S w_s)}{(\lambda\sigma(\bm{w}))^2}<0,$$
where the inequality follows from the fact that $\sum_{s=1}^S w_s> 0$, and that as $w_s<p_s\forall s$ and $\lambda \leq \underset{s\in \{1,...,S\}}{\min}p_s/w_s$, $1-\lambda \sum_{s=1}^S w_s\geq 0$. Therefore, for any $\lambda \in \left(1,\underset{s\in \{1,...,S\}}{\min}p_s/w_s\right]$,
\begin{equation}\label{eq:conditions_under_whichpowerlower_proof1}
\frac{\overline{|B|}(\lambda\bm{w})}{\sigma(\lambda\bm{w})}< \frac{\overline{|B|}(\bm{w})}{\sigma(\bm{w})}.
\end{equation}
Now,
\begin{align*}
&0\in \text{CI}_{1-\alpha}(\lambda\bm{w})\\
\Leftrightarrow~&0\in \left[\widehat{\tau}(\lambda\bm{w})-Q_{1-\alpha}\left(\overline{|B|}(\lambda\bm{w}),\sigma(\lambda\bm{w})\right),\widehat{\tau}(\lambda\bm{w})+Q_{1-\alpha}\left(\overline{|B|}(\lambda\bm{w}),\sigma(\lambda\bm{w})\right)\right]\\
\Leftrightarrow~&0\in \left[\lambda\widehat{\tau}(\bm{w})-\lambda\sigma(\bm{w})Q_{1-\alpha}\left(\overline{|B|}(\lambda\bm{w})/\sigma(\lambda\bm{w}),1\right),\lambda\widehat{\tau}(\bm{w})+\lambda\sigma(\bm{w})Q_{1-\alpha}\left(\overline{|B|}(\lambda\bm{w})/\sigma(\lambda\bm{w}),1\right)\right]\\
\Leftrightarrow~&0\in \left[\widehat{\tau}(\bm{w})-\sigma(\bm{w})Q_{1-\alpha}\left(\overline{|B|}(\lambda\bm{w})/\sigma(\lambda\bm{w}),1\right),\widehat{\tau}(\bm{w})+\sigma(\bm{w})Q_{1-\alpha}\left(\overline{|B|}(\lambda\bm{w})/\sigma(\lambda\bm{w}),1\right)\right]\\
\Rightarrow~&0\in \left[\widehat{\tau}(\bm{w})-\sigma(\bm{w})Q_{1-\alpha}\left(\overline{|B|}(\bm{w})/\sigma(\bm{w}),1\right),\widehat{\tau}(\bm{w})+\sigma(\bm{w})Q_{1-\alpha}\left(\overline{|B|}(\bm{w})/\sigma(\bm{w}),1\right)\right]\\
\Leftrightarrow~&0\in \text{CI}_{1-\alpha}(\bm{w}).
\end{align*}
The second equivalence follows from the fact that $Q_{1-\alpha}\left(\mu,\sigma\right)=\sigma Q_{1-\alpha}\left(\mu/\sigma,1\right)$ and $\widehat{\tau}(\lambda\bm{w})=\lambda\widehat{\tau}(\bm{w})$ and $\sigma(\lambda\bm{w})=\lambda\sigma(\bm{w}).$ The third equivalence follows from $\lambda>0$. The implication follows from \eqref{eq:conditions_under_whichpowerlower_proof1} and the fact that
$\mu\mapsto Q_{1-\alpha}\left(\mu,1\right)$ is increasing on $\mathbb{R}_+$. The last equivalence follows from the fact that
$\sigma Q_{1-\alpha}\left(\mu/\sigma,1\right)=Q_{1-\alpha}\left(\mu,\sigma\right).$

\subsection{Proof of Proposition \ref{prop:conditions_under_whichpowerlower2}}

Let $\Phi$ be the cdf of the standard normal distribution.
\begin{align}\label{eq:power}
&P(0\in \text{CI}_{1-\alpha}(\bm{w}))\nonumber\\
=&P\left(0\in \left[\widehat{\tau}(\bm{w})-Q_{1-\alpha}\left(\overline{|B|}(\bm{w}),\sigma(\bm{w})\right),\widehat{\tau}(\bm{w})+Q_{1-\alpha}\left(\overline{|B|}(\bm{w}),\sigma(\bm{w})\right)\right]\right)\nonumber\\
=&P\left(-Q_{1-\alpha}\left(\overline{|B|}(\bm{w}),\sigma(\bm{w})\right)\leq \widehat{\tau}(\bm{w})\leq Q_{1-\alpha}\left(\overline{|B|}(\bm{w}),\sigma(\bm{w})\right)\right)\nonumber\\
=&P\left(-\frac{E\left(\widehat{\tau}(\bm{w})\right)}{\sigma(\bm{w})}-Q_{1-\alpha}\left(\overline{|B|}(\bm{w})/\sigma(\bm{w}),1\right)\leq \frac{\widehat{\tau}(\bm{w})-E\left(\widehat{\tau}(\bm{w})\right)}{\sigma(\bm{w})} \leq -\frac{E\left(\widehat{\tau}(\bm{w})\right)}{\sigma(\bm{w})}+Q_{1-\alpha}\left(\overline{|B|}(\bm{w})/\sigma(\bm{w}),1\right)\right)\nonumber\\
=&\Phi\left(-\frac{E\left(\widehat{\tau}(\bm{w})\right)}{\sigma(\bm{w})}+Q_{1-\alpha}\left(\overline{|B|}(\bm{w})/\sigma(\bm{w}),1\right)\right)-\Phi\left(-\frac{E\left(\widehat{\tau}(\bm{w})\right)}{\sigma(\bm{w})}-Q_{1-\alpha}\left(\overline{|B|}(\bm{w})/\sigma(\bm{w}),1\right)\right)\\
=&\Phi\left(-\frac{\sum_{s=1}^S w_s\tau_s}{\sigma(\bm{w})}+Q_{1-\alpha}\left(\overline{|B|}(\bm{w})/\sigma(\bm{w}),1\right)\right)-\Phi\left(-\frac{\sum_{s=1}^S w_s\tau_s}{\sigma(\bm{w})}-Q_{1-\alpha}\left(\overline{|B|}(\bm{w})/\sigma(\bm{w}),1\right)\right)\nonumber\\
=&\Phi\left(-\frac{\tau \sum_{s=1}^S w_s}{\sigma(\bm{w})}+Q_{1-\alpha}\left(\overline{|B|}(\bm{w})/\sigma(\bm{w}),1\right)\right)-\Phi\left(-\frac{\tau \sum_{s=1}^S w_s}{\sigma(\bm{w})}-Q_{1-\alpha}\left(\overline{|B|}(\bm{w})/\sigma(\bm{w}),1\right)\right)\nonumber\\
=:&\mathcal{P}\left(\tau,\frac{\sum_{s=1}^S w_s}{\sigma(\bm{w})},Q_{1-\alpha}\left(\overline{|B|}(\bm{w})/\sigma(\bm{w}),1\right)\right).\nonumber
\end{align}
The third equality follows from the fact that $Q_{1-\alpha}\left(\mu,\sigma\right)=\sigma Q_{1-\alpha}\left(\mu/\sigma,1\right)$, the fourth follows from Assumption \ref{hyp:CATE_hat_normal}, the fifth follows from Assumption \ref{hyp:CATE_hat}, and the sixth follows from $\sum_{s=1}^S p_s \frac{w_s}{p_s}\tau_s=\sum_{s=1}^S p_s \frac{w_s}{p_s}\times \sum_{s=1}^S p_s \tau_s$.
$\frac{\sum_{s=1}^S w_s}{\sigma(\bm{w})}>0$, $Q_{1-\alpha}\left(\overline{|B|}(\bm{w})/\sigma(\bm{w}),1\right)>0$, $\frac{\sum_{s=1}^S p_s}{\sigma(\bm{p})}=\frac{1}{\sigma(\bm{p})}>0$, and $Q_{1-\alpha}\left(\overline{|B|}(\bm{p})/\sigma(\bm{p}),1\right)>0$. For $x>0$ and $y>0$, letting $\phi$ denote the pdf of the standard normal,
\begin{align*}
\frac{\partial \mathcal{P}\left(w,x,y\right)}{\partial x}=w(\phi(-wx-y)-\phi(-wx+y))\leq 0.
\end{align*}
If $w>0$ (resp. $w<0$), the inequality follows because
$|-wx-y|>|-wx+y|$ (resp. $|-wx-y|<|-wx+y|$) as $u\mapsto\phi(u)$ is symmetric and decreasing in $|u|$.
\begin{align*}
\frac{\partial \mathcal{P}\left(w,x,y\right)}{\partial y}=\phi(-wx-y)+\phi(-wx+y)>0.
\end{align*}
Then, if
$\frac{\sum_{s=1}^S w_s}{\sigma(\bm{w})}<\frac{\sum_{s=1}^S p_s}{\sigma(\bm{p})}$, as
$\overline{|B|}(\bm{w})/\sigma(\bm{w})> \overline{|B|}(\bm{p})/\sigma(\bm{p})=0$ and $\mu\mapsto Q_{1-\alpha}\left(\mu,1\right)$ is strictly increasing on $\mathbb{R}_+$, the result holds since $\mathcal{P}\left(w,x,y\right)$ is decreasing in $x$ and strictly increasing in $y$.

\section*{Auxiliary Lemmas}

\begin{lem}\label{lem_decreasingthrehsold}
If $\underline{s}<S$, for any $s \in \{\underline{s},...,S-1\}$,
$$\frac{1}{\frac{1}{B^2}+\sum_{s'=s}^S\frac{1}{V_{s'}}}\sum_{s'=s}^Sp_{s'}\geq \frac{1}{\frac{1}{B^2}+\sum_{s'=s+1}^S\frac{1}{V_{s'}}}\sum_{s'=s+1}^Sp_{s'}.$$
\end{lem}
\textbf{Proof of Lemma \ref{lem_decreasingthrehsold}:}\\
\begin{align*}
&\frac{1}{\frac{1}{B^2}+\sum_{s'=s}^S\frac{1}{V_{s'}}}\sum_{s'=s}^Sp_{s'}\geq \frac{1}{\frac{1}{B^2}+\sum_{s'=s+1}^S\frac{1}{V_{s'}}}\sum_{s'=s+1}^Sp_{s'}\\
\Leftrightarrow & \sum_{s'=s}^Sp_{s'}\left( \frac{1}{B^2}+\sum_{s'=s+1}^S\frac{1}{V_{s'}}\right)\geq \sum_{s'=s+1}^Sp_{s'}\left(\frac{1}{B^2}+\sum_{s'=s}^S\frac{1}{V_{s'}}\right)\\
\Leftrightarrow & p_{s}\left( \frac{1}{B^2}+\sum_{s'=s+1}^S\frac{1}{V_{s'}}\right)\geq \sum_{s'=s+1}^Sp_{s'}\times \frac{1}{V_{s}}\\
\Leftrightarrow & p_{s}V_{s}\geq \frac{1}{\frac{1}{B^2}+\sum_{s'=s+1}^S\frac{1}{V_{s'}}} \sum_{s'=s+1}^Sp_{s'}.
\end{align*}
The result follows from the previous display and Lemma \ref{lem_smaller_than_higher_thresholds} \textbf{QED.}

\begin{lem}\label{lem_smaller_than_higher_thresholds}
If $\underline{s}<S$, for any $s \in \{\underline{s},...,S-1\}$,
$$p_{s}V_{s}\geq \frac{1}{\frac{1}{B^2}+\sum_{s'=s+1}^S\frac{1}{V_{s'}}} \sum_{s'=s+1}^Sp_{s'}.$$
\end{lem}

\textbf{Proof of Lemma \ref{lem_smaller_than_higher_thresholds}:}\\

By Lemma \ref{lem:includedset},
$$p_s V_s\geq \frac{1}{\frac{1}{B^2}+\sum_{s'=s}^S\frac{1}{V_{s'}}} \sum_{s'=s}^Sp_{s'}.$$
Then,
\begin{align*}
&p_{s} V_{s}\left(\frac{1}{B^2}+\sum_{s'=s+1}^S\frac{1}{V_{s'}}\right)\\
=&p_{s} V_{s}\left(\frac{1}{B^2}+\sum_{s'=s}^S\frac{1}{V_{s'}}\right)-p_s\\
\geq &\sum_{s'=s}^Sp_{s'}-p_{s}\\
=&\sum_{s'=s+1}^Sp_{s'}.
\end{align*}
\textbf{QED.}

\begin{lem}\label{lem:includedset}
\begin{align*}
&p_s V_s \geq \frac{1}{\frac{1}{B^2}+\sum_{s'=s}^S\frac{1}{V_{s'}}} \sum_{s'=s}^Sp_{s'} \Rightarrow  p_{s+1} V_{s+1} \geq \frac{1}{\frac{1}{B^2}+\sum_{s'=s+1}^S\frac{1}{V_{s'}}} \sum_{s'=s+1}^Sp_{s'}.
\end{align*}
\end{lem}

\textbf{Proof of Lemma \ref{lem:includedset}:}\\

Assume that
$$p_s V_s\geq \frac{1}{\frac{1}{B^2}+\sum_{s'=s}^S\frac{1}{V_{s'}}} \sum_{s'=s}^Sp_{s'}.$$
Then,
\begin{align*}
&p_{s+1} V_{s+1}\left(\frac{1}{B^2}+\sum_{s'=s+1}^S\frac{1}{V_{s'}}\right)\\
=&p_{s+1} V_{s+1}\left(\frac{1}{B^2}+\sum_{s'=s}^S\frac{1}{V_{s'}}\right)-p_{s+1}\frac{V_{s+1}}{V_{s}}\\
=&p_{s} V_{s}\left(\frac{1}{B^2}+\sum_{s'=s}^S\frac{1}{V_{s'}}\right)+\left(p_{s+1} V_{s+1}-p_{s} V_{s}\right)\sum_{s'=s}^S\frac{1}{V_{s'}}+\left(p_{s+1} V_{s+1}-p_{s} V_{s}\right)\frac{1}{B^2}-p_{s+1}\frac{V_{s+1}}{V_{s}}\\
\geq &p_{s} V_{s}\left(\frac{1}{B^2}+\sum_{s'=s}^S\frac{1}{V_{s'}}\right)+p_{s+1}\frac{V_{s+1}}{V_{s}}-p_{s}-p_{s+1}\frac{V_{s+1}}{V_{s}}\\
\geq &\sum_{s'=s+1}^Sp_{s'}.
\end{align*}
The first inequality follows from $p_{s+1} V_{s+1}-p_{s} V_{s}\geq 0$, the second follows from the assumption in the first display of the proof.
\textbf{QED.}

\medskip
\textbf{QED.}

\begin{lem}\label{lem:positive_mus}
Assume that $B\geq 1$. If $\frac{V_S}{1/S\sum_{s'=1}^{S}V_{s'}}\leq B+1$, then for all $s\in \{1,...,S\}$
$\sum_{s'=1}^{s}V_{s'}/\sum_{s'=1}^{S}V_{s'}\geq s(B+1)/S-B.$
\end{lem}

\medskip
\textbf{Proof of Lemma \ref{lem:positive_mus}}

Let $U_s=\sum_{s'=1}^{s}V_{s'}/\sum_{s'=1}^{S}V_{s'}+B(1-s/S)-s/S$.
$U_S=1+B(1-1)-1=0$.
Now,
$$U_{s+1}-U_{s}=V_{s+1}/\sum_{s'=1}^{S}V_{s'}-(B+1)/S\leq V_{S}/\sum_{s'=1}^{S}V_{s'}-(B+1)/S \leq 0 $$
by assumption. Therefore, for all $s$
$U_s\geq U_S=0.$

\medskip
\textbf{QED.}

\end{document}